\documentclass{emulateapj}

\shorttitle{HST images of post-AGB objects}
\shortauthors{Si\'odmiak et al.}

\begin{document}

\title{HST Snapshot Survey of Post-AGB Objects}


\author{N. Si\'odmiak}
\affil{Space Telescope Science Institute, 3700 San Martin Drive, Baltimore, MD 21218, USA}
\affil{N. Copernicus Astronomical Center, Rabia\'nska 8, 87-100 Toru\'n, Poland}
 \email{siodmiak@stsci.edu, siodmiak@ncac.torun.pl}

\author{M. Meixner}
\affil{Space Telescope Science Institute, 3700 San Martin Drive, Baltimore, MD 21218, USA}
\email{meixner@stsci.edu}

\author{T. Ueta}
\affil{Department of Physics and Astronomy, University of Denver, 2112 E. Wesley Avenue, Denver, CO 80208, USA}
\email{tueta@du.edu}

\author{B.E.K. Sugerman}
\affil{Goucher College, 1021 Dulaney Valley Road, Baltimore, MD 21204, USA}
\email{ben.sugerman@goucher.edu}

\author{G.C. Van de Steene}
\affil{Royal Observatory of Belgium, Ringlaan 3, 1180 Brussels, Belgium }
\email{gsteene@oma.be}

\and

\author{R. Szczerba}
\affil{N. Copernicus Astronomical Center, Rabia\'nska 8, 87-100 Toru\'n, Poland}
\email{szczerba@ncac.torun.pl}

\begin{abstract}
The results from a $Hubble$ $Space$ $Telescope$ ($HST$) snapshot survey of
post-AGB objects are shown. The aim of the survey is to complement existing
$HST$ images of PPN and to connect various types of nebulosities with
physical and chemical properties of their central stars. Nebulosities are
detected in 15 of 33 sources. Images and photometric and geometric
measurements are presented. For sources with nebulosities we see a
morphological bifurcation into two groups, DUPLEX and SOLE, as previous
studies have found. We find further support to the previous results
suggesting that this dichotomy is caused by a difference in optical
thickness of the dust shell. The remaining 18 sources are classified as
stellar post-AGB objects, because our observations indicate a lack of
nebulosity. We show that some stellar sources may in fact be DUPLEX or SOLE
based on their infrared colors. The cause of the differences among the
groups are investigated. We discuss some evidence suggesting that high
progenitor-mass AGB stars tend to become DUPLEX post-AGB objects.
Intermediate progenitor-mass AGB stars tend to be SOLE post-AGB objects.
Most of the stellar sources probably have low mass progenitors and do not
seem to develop nebulosities during the post-AGB phase and therefore do not 
become planetary nebulae.
\end{abstract}

\keywords{planetary nebulae: general --- stars: AGB and post-AGB --- stars:
circumstellar matter --- stars: mass loss --- reflection nebulae}


\section{Introduction}
The post-AGB\footnote{The term ``proto-planetary nebulae'' (PPN) is also
often used to describe objects in transition between AGB and PN. However,
one has to remember that low-mass objects that evolve very slowly will not
be able to ionize the ejected matter and become planetary nebulae. Hence,
the term ``post-AGB'' describes a wider group of evolved objects.} phase is
a short period in the evolution of low- and intermediate-mass stars ($0.8 -
8M_{\odot}$) between the asymptotic giant branch (AGB) and the planetary
nebula (PN) phases. The theoretical evolutionary tracks of the central stars
\citep{blo95} predict that this phase lasts $10^2 - 10^5$ yrs, depending on 
the core mass, yet even shorter kinematical age for nebulosities, of order 
of $10^3 - 10^4$ yrs \citep{sah07}. Shorter kinematical life time is obvious
from the observational point of view since the entire shell (especially an
extended cold shell of post-AGB object) is not necessarily detected at the
optical wavelength.

Many significant changes occur during this phase to both the star and the
nebula. The superwind (suddenly increased mass loss at the end of the AGB
phase) stops, large amplitude pulsations cease and the circumstellar
envelope slowly expands and cools revealing the central star. The post-AGB
star evolves at constant luminosity on the Hertzsprung-Russell diagram
towards higher temperatures. The morphology of the expanding envelope
changes, resulting in the fascinating shapes of the nebula itself. 

Despite many efforts the post-AGB phase is still poorly understood.
Especially the departure from spherically symmetric AGB circumstellar
envelopes \citep[e.g.,][]{olo01} and the formation of diverse, axisymmetric,
nebulae \citep[e.g.,][]{bal02} are of great interest to astronomers.
Determining the main cause(s) of the breaking of mass-loss symmetry is an
important issue in understanding the late stages of stellar evolution.

Previous morphological studies of post-AGB objects \citep[e.g.,][hereafter
UMB00]{sah99,kwok00,hri00,hri01,ueta00} revealed many asymmetrical nebulae
around central stars, where jets and concentric arcs or rings were not
unusual. Further, the ground-based mid-IR (8--21 $\mu$m) observations done
by \citet{mei99} showed two morphological classes of extended axisymmetric
nebulae among post-AGB objects: ``core/elliptical'' with unresolved cores
and elliptical nebulae and ``toroidal'' with limb-brightened peaks that
suggest equatorial density enhancements. 

Optical observations revealed more interesting details on structures seen in
post-AGB nebulae. UMB00 observed 27 PPN candidates and divided them into two
classes: the Star-Obvious Low-level Elongated (SOLE) nebulae which show
star-dominated emission with faint extended nebulosity and DUst-Prominent
Longitudinally EXtended (DUPLEX) nebulae which are dust dominated with a
faint or completely obscured central star.

They also noticed that those two optical classes corresponded to the mid-IR
ones, i.e. core/elliptical $\approx$ DUPLEX and toroidal $\approx$ SOLE, and
that the main difference between SOLE and DUPLEX nebulae was the degree of
the equatorial enhancement (i.e., DUPLEXes tend to have a higher
equator-to-pole density ratio than SOLEs) and could not be attributed only
to the inclination-angle effects. They also suggested that SOLE nebulae were
optically thinner than DUPLEX nebulae and that SOLE PPNs would have evolved
from low-mass progenitors and DUPLEX from high-mass progenitors. Later model
calculations \citep{mei02} confirmed the physical differences between those
two classes of objects. Other high-resolution near-infrared $HST$
observations \citep[e.g.,][]{ueta05,su03} also supported optical results.
Follow-up studies by e.g., \citet{ueta05} were conducted to further
investigate the asymmetry of observed objects. They also found that
differences in observed morphologies are caused (mainly) by optical depth
and (to a lesser extent) by inclination of the objects. All those
observations strengthen the hypothesis of an equatorial mass loss
enhancement at the end of the AGB (superwind) phase \citep{mei99} and that
morphological shaping of nebulae occurs before the post-AGB phase. Moreover,
the resulting diverse shapes observed in proto-planetary as well as PNs may
be caused by some properties of the central stars such as chemical
composition, mass and/or metallicity.

In order to find further support to those results we have executed an $HST$
snapshot survey of post-AGB objects by increasing the range of masses and
chemical composition (C/O ratio) and by enlarging the sample of studied
objects. The goal of the project is to connect the observed diversity of
nebular shapes with physical and chemical properties of the central stars.
In this paper we present the results for post-AGB objects observed recently
with $HST$. Observation and data reduction are summarized in $\S$2. The
results and differences between selected groups of objects are shown in
$\S$3. In $\S$4 we discuss obtained results with references to physical and
chemical properties of stars. The conclusions are included in $\S$5.

\begin{table*}[!ht]
\begin{center}
{Table 1: Target summary for $HST$ ACS/HRC observation of post-AGB objects}
\small
\begin{tabular}{llrrrrrl}
\tableline\tableline \multicolumn{1}{c}{IRAS ID} & \multicolumn{1}{c}{Other
name} & \multicolumn{1}{c}{Prop. ID$^{1}$} & \multicolumn{1}{c}{$V$
mag$^{2}$} & \multicolumn{1}{c}{J mag$^{2}$} & \multicolumn{1}{c}{T$_{\rm
eff}$$^{3,4}$} &
\multicolumn{1}{c}{Mass$^4$} & \multicolumn{1}{c}{C/O$^4$} \\
\tableline
01005+7910       & ...         & 10627~~~~ & 10.96~~ & 10.274~ &21000~~    &     0.55~~~~ & ~~C     \\
04395+3601       & AFGL 618    &  9430~~~~ & ...~~~~ & 13.510~ &25000\tablenotemark{a}~&    ...~~~~~~ &~~C\tablenotemark{a} \\
06034+1354       & DY Ori      & 10627~~~~ & ...~~~~ &  8.073~ & 5900~~    & $<$ 0.55~~~~ & ~~O~    \\
08143$-$4406     & PM 1$-$39   & 10627~~~~ & ...~~~~ &  9.172~ & 7150~~    & $<$ 0.55~~~~ & ~~C~    \\
09256$-$6324     & IW Car      & 10627~~~~ &  8.33~~ &  5.875~ & 6700~~    & $<$ 0.55~~~~ & ~~C~    \\
11385$-$5517$^5$ & V885 Cen    &  9436~~~~ &  7.04~~ &  5.947~ & 8500~~    &     0.55~~~~ & ~~O~    \\
12067$-$4508     & RU Cen      & 10627~~~~ &  9.13~~ &  7.616~ & 6000~~    & $<$ 0.55~~~~ & ~~O~    \\
12175$-$5338     & V1024 Cen   & 10627~~~~ &  9.40~~ &  8.326~ & 7350~~    &     0.62~~~~ & ~~O~    \\
12222$-$4652     & HD 108015   & 10627~~~~ &  8.01~~ &  6.941~ & 6800~~    & $<$ 0.55~~~~ & ~~O~    \\
12538$-$2611     & LN Hya      & 10627~~~~ &  6.82~~ &  5.251~ & 6000~~    & $<$ 0.55~~~~ & ~~O~    \\
13416$-$6243     & ...         & 10627~~~~ & ...~~~~ & 10.302~ &...~~~~    &    ...~~~~~~ & ~~C\tablenotemark{b} \\
13428$-$6232$^5$ & ...         &  9463~~~~ & ...~~~~ & 13.176~ &...~~~~    &    ...~~~~~~ & ~~C\tablenotemark{b} \\
15039$-$4806     & HD 133656   & 10627~~~~ &  7.58~~ &  6.680~ & 8000~~    &     0.55~~~~ & ~~O~    \\
15469$-$5311     & ...         & 10627~~~~ & 10.82~~ &  7.190~ & 7500~~    & $<$ 0.55~~~~ & ~~O~    \\
15553$-$5230$^5$ & ...         & 10627~~~~ & ...~~~~ & 13.380~ &...~~~~    &    ...~~~~~~ & ~~...   \\
16206$-$5956     & LS 3591     & 10627~~~~ & 10.00~~ &  9.002~ & 8500~~    &     0.66~~~~ & ~~O~    \\
17163$-$3907     & Hen 3$-$1379& 10185~~~~ & 12.45~~ &  4.635~ &...~~~~    &    ...~~~~~~ & ~~...   \\
17243$-$4348     & LR Sco      & 10627~~~~ & 10.49~~ &  8.035~ & 6750~~    &     0.94~~~~ & ~~O~    \\
17279$-$1119     & V340 Ser    & 10627~~~~ &  9.78~~ &  7.845~ & 7300~~    & $<$ 0.55~~~~ & ~~C~    \\
17516$-$2525     & ...         &  9436~~~~ & 17.76~~ &  8.695~ &...~~~~    &    ...~~~~~~ & ~~O\tablenotemark{b} \\
17534+2603       & 89 Her      & 10627~~~~ &  5.51~~ &  4.998~ & 6550~~    &     0.61~~~~ & ~~O~    \\
18135$-$1456     & OH 15.7+0.8 &  9436~~~~ & 16.61~~ & ...~~~~ &...~~~~    &    ...~~~~~~ & ~~O\tablenotemark{c} \\
19125+0343       & BD+03 3950  & 10627~~~~ & 10.46~~ &  7.903~ & 7750~~    &     0.58~~~~ & ~~O~    \\
19157$-$0247     & BD$-$02 4931& 10627~~~~ & 10.87~~ &  8.872~ & 7750~~    &     0.58~~~~ & ~~O~    \\
19306+1407$^5$   & ...         &  9436~~~~ & ...~~~~ & 11.286~ & B~~~~     &    ...~~~~~~ & ~~O\tablenotemark{b} \\
19475+3119$^5$   & HD 331319   &  9436~~~~ &  9.60~~ &  7.773~ & 7750~~    &     0.58~~~~ & ~~O~    \\
20000+3239$^5$   & ...         &  9436~~~~ & ...~~~~ &  8.021~ & 5500\tablenotemark{d}~&    ...~~~~~~ & ~~C\tablenotemark{b} \\
20117+1634       & R Sge       & 10627~~~~ &  9.31~~ &  7.818~ & 5000~~    &     0.93~~~~ & ~~O~    \\
20547+0247       & U Equ       &  9436~~~~ & ...~~~~ & 11.561~ & G~~~~     &    ...~~~~~~ & ~~O\tablenotemark{e} \\
22036+5306$^5$   & ...         & 10185~~~~ & ...~~~~ & 11.666~ &...~~~~    &    ...~~~~~~ & ~~O\tablenotemark{b} \\
22223+4327$^5$   & BD+42 4388  &  9436~~~~ & 10.00~~ &  7.812~ & 6500~~    &     0.55~~~~ & ~~C~    \\
23304+6147$^5$   & PM 2$-$47   &  9436~~~~ & ...~~~~ &  8.501~ & 6750~~    &     0.66~~~~ & ~~C~    \\
23541+7031       & M 2$-$56    &  9436~~~~ & ...~~~~ & 13.857~ & B~~~~     &    ...~~~~~~ & ~~O\tablenotemark{b} \\
\tableline
\end{tabular}
\end{center}
\small
$^1$ 9430 - PI S. Trammell, 9436 \& 10185 - PI R.Sahai, 10627 - PI M. Meixner.\\
$^2$ Magnitudes from GSC2.2 and 2MASS catalogues.\\
$^3$ Spectral type is shown if T$_{\rm eff}$ is not known from the model atmosphere analysis.\\
$^4$ Data collected by \citet{sta06} with few exceptions:
$^a$ \citet{cer01} and references therein,
$^b$ C/O ratio established by the inspection of the ISO spectrum,
$^c$ \citet{eng02} and references therein,
$^d$ \citet{volk02},
$^e$ \citet{geb05} and references therein.\\
$^5$ Images of those objects were also published in the newest paper by
\citet{sah07}.
\vspace{0.5cm}
\end{table*}

\section{Observations}
\subsection{Sample selection}
Post-AGB stars do not form a homogeneous group. One can find objects with
different masses (including very low-mass objects that will never become
PNs) and chemical composition. However, they are expected to be supergiants
of spectral type B--K with an evidence of the circumstellar envelope (e.g.,
infrared excess) and no large photometric variability \citep[the
classification criteria for an object to be a post-AGB star were described
in e.g.,][]{kwok93,win03,wae04}. In addition, few unique classes of stars
like RV Tau objects \citep{jura86}, R CrB stars and extreme helium stars
\citep{ren79,iben83} or even ``born-again'' supergiants \citep[Sakurai's
object and FG Sge, e.g.,][]{law03} are considered as evolved objects in the
post-AGB phase. All those objects are included in the Toru\'n catalogue of
galactic post-AGB and related objects
\citep[][http://www.ncac.torun.pl/postagb]{sz07}. From this database we
selected for our snapshot survey\footnote{\,Snapshot surveys are designed 
to fill the observational schedule gaps between larger GO programs and there 
is no guarantee that any particular snapshot target will be observed. That 
is why we may expect only part of the selected sample to be observed.} those
sources that were not yet imaged with $HST$. Because the exposure time for
an object from a snapshot survey cannot be too long, our sample was biased
towards rather bright stars.

The selected post-AGB objects were observed with the High Resolution Channel
(HRC) of the Advanced Camera for Surveys (ACS) on-board $HST$ \citep{gon05},
which has a $26\arcsec$ x $29\arcsec$ field of view and a plate-scale of
0.027$\arcsec$ pixel$^{-1}$. Observations for 19 objects from our proposal
(program ID 10627, PI M. Meixner) were done between July 2005 and January
2007. We also searched for other post-AGB stars (from the Toru\'n catalogue)
observed with ACS/HRC. In the $HST$ Archive we found 17 objects from
previous observation cycles. We included 14 post-AGB sources in our
reduction and analysis of the images (selected objects from programs ID 9463
\& 10185, PI R. Sahai and 9430, PI S. Trammell) and excluded 3 objects (the
extended structures of Egg Nebula and Frosty Leo are bigger than the size of
an image what prevents the correct measurements and IRAS 19024+0044 was
already carefully analyzed by \citet{sah05}). Together we have 33 PPNs
observed in broad $B$ (F435W), $V$ (F606W) and $I$ (F814W) filters. Selected
candidates differ in central star masses, optical and infrared colors and
effective temperatures \citep[and references therein]{sta06}. Thereby we
increased the number of studied post-AGB objects and covered more
extensively the diverse nature of post-AGB stars. The properties of the
analyzed post-AGB objects are shown in Table 1.

\begin{table*}[ht]
\begin{center}
{Table 2: Properties of SOLE post-AGB objects}
\small
\begin{tabular}{lrrllrrrr}
\tableline\tableline
IRAS ID & \multicolumn{2}{c}{Obs. Coord. (J2000)} & \multicolumn{1}{c}{ACS} & \multicolumn{1}{c}{$F_{\lambda}^{~a}$} &
\multicolumn{1}{c}{HST} & $I_{s}/I_{n}^{~c}$ & \multicolumn{1}{c}{Size} & \multicolumn{1}{c}{$e^{~d}$}\\
 & \multicolumn{1}{c}{ra} & \multicolumn{1}{c}{dec} & \multicolumn{1}{c}{filter} & & \multicolumn{1}{c}{Mag$^{b}$}&&
\multicolumn{1}{c}{(arcsec) ($\sigma$)}&\\
\tableline
01005+7910  &01 04 45.59&   79 26 47.08&F606W&1.23e$-$13&11.17~~& 26900& 3.75 x 2.20 (\,~2)&0.41\\
            &           &              &F814W&5.94e$-$14&11.97~~& 14510& 3.65 x 2.25 (\,~4)&0.38\\
08143$-$4406&08 16 02.98&$-$44 16 04.88&F606W&5.58e$-$14&12.03~~&  7500& 2.10 x 1.30 (10)  &0.38\\
            &           &              &F814W&7.61e$-$14&11.70~~&  3700& 2.10 x 1.35 (\,~4)&0.36\\
11385$-$5517&11 40 58.82&$-$55 34 25.52&F435W&8.10e$-$12& 6.63~~&338000& 8.30 x 6.60 (12)  &0.20\\
            &           &              &F606W&4.47e$-$12$^e$&7.27$^e$& 81500& 8.30 x 6.20 (\,~8)&0.25\\
12175$-$5338&12 20 15.08&$-$53 55 31.62&F606W&6.46e$-$13& 9.37~~& 76000& 5.70 x 2.50 (\,~7)&0.56\\
            &           &              &F814W&3.76e$-$13& 9.96~~& 36300& 5.60 x 2.20 (10)  &0.61\\
16206$-$5956&16 25 02.61&$-$60 03 31.35&F606W&3.33e$-$13&10.09~~& 21500& 3.85 x 2.60 (\,~9)&0.32\\
            &           &              &F814W&1.87e$-$13&10.72~~& 21300& 3.85 x 2.55 (\,~6)&0.34\\
19306+1407  &19 32 55.15&  +14 13 36.89&F606W&7.83e$-$15&14.17~~&  9000& 7.70 x 2.00 (\,~4)&0.74\\
            &           &              &F814W&1.09e$-$14&13.81~~&  6400& 4.40 x 2.05 (\,~4)&0.53\\
19475+3119  &19 49 29.63&  +31 27 15.32&F435W&6.91e$-$13& 9.30~~&190000&10.45 x 5.10 (\,~3)&0.51\\
            &           &              &F606W&6.72e$-$13& 9.33~~&153000& 9.95 x 5.10 (\,~2)&0.49\\
20000+3239  &20 01 59.56&  +32 47 33.71&F606W&3.39e$-$14&12.57~~&  3800& 2.15 x 1.45 (\,~5)&0.33\\
            &           &              &F814W&1.02e$-$13&11.38~~&  2000& 2.00 x 1.20 (\,~7)&0.40\\
22223+4327  &22 24 31.48&  +43 43 10.67&F435W&3.47e$-$13&10.05~~& 13000& 3.40 x 2.10 (10)  &0.38\\
            &           &              &F606W&5.02e$-$13& 9.65~~& 10000& 3.50 x 2.10 (\,~6)&0.40\\
23304+6147  &23 32 44.71&  +62 03 48.92&F606W&3.30e$-$14&12.61~~&  3800& 2.25 x 1.60 (14)  &0.29\\
            &           &              &F814W&8.31e$-$14&11.60~~&  2100& 2.20 x 1.45 (\,~9)&0.34\\
\tableline
\end{tabular}
\end{center}
\small
$^a$ In units of erg s$^{-1}$ cm$^{-2}$ \AA$^{-1}$.\\
$^b$ HST system magnitudes (STMAG).\\
$^c$ Star surface intensity $I_{s}$ to nebula surface intensity $I_{n}$ ratio.\\
$^d$ Ellipticity $e = 1 - b/a$, where $a$ and $b$ are major- and minor-axis lengths, respectively.\\
$^e$ Due to a saturation problem, $F_{\lambda}$ and HST Mag are only the lower limits.
\vspace{0.3cm}
\end{table*}

\begin{table*}[!ht]
\vspace{0.35cm}
\begin{center}
{Table 3: Properties of DUPLEX post-AGB objects}
\small
\begin{tabular}{lrrlrrrrr}
\tableline\tableline
IRAS ID & \multicolumn{2}{c}{Obs. Coord. (J2000)} & \multicolumn{1}{c}{ACS} & \multicolumn{1}{c}{$F_{\lambda}^{~a}$} &
\multicolumn{1}{c}{HST} & $I_{s}/I_{n}^{~c}$ & \multicolumn{1}{c}{Size} & \multicolumn{1}{c}{$e^{~d}$}\\
 & \multicolumn{1}{c}{ra} & \multicolumn{1}{c}{dec} & \multicolumn{1}{c}{filter} & & \multicolumn{1}{c}{Mag$^{b}$}&&
\multicolumn{1}{c}{(arcsec) ($\sigma$)}&\\
\tableline
04395+3601  &04 42 53.50&   +36 06 51.71~~   &F606W&2.54e$-$15&15.39&...&15.40 x 4.80 (4)&0.67\\
            &\multicolumn{2}{c}{eastern lobe}&     &1.98e$-$15&15.66&124& 8.25 x 4.75 (4)&0.42\\
            &\multicolumn{2}{c}{western lobe}&     &5.16e$-$16&17.12& 85& 6.20 x 3.20 (4)&0.48\\
13428$-$6232&13 46 20.96& $-$62 47 58.10$^e$ &F606W&...~~~~~&...~~~&...~&...~~~~~~~~~~~~&...~~\\
            &           &                    &F814W&3.86e$-$16&17.43&  4& 8.75 x 6.90 (1)&0.21\\
15553$-$5230&15 59 10.66&$-$52 38 38.15~~    &F814W&1.90e$-$17&20.70&...& 2.50 x 1.10 (2)&0.56\\
            &\multicolumn{2}{c}{eastern lobe}&     &1.21e$-$17&21.20& 10& 1.35 x 1.00 (2)&0.26\\
            &\multicolumn{2}{c}{western lobe}&     &6.01e$-$18&21.95&  4& 1.20 x 0.85 (2)&0.29\\
22036+5306  &22 05 30.35&   +53 21 33.97~~   &F606W&1.32e$-$15&16.10&330& 6.75 x 2.30 (3)&0.66\\
            &           &                    &F814W&2.87e$-$15&15.26&120& 5.60 x 1.85 (4)&0.67\\
23541+7031  &23 56 36.31&   +70 48 18.39~~   &F606W&5.33e$-$17&19.58& 55& 3.75 x 1.60 (2)&0.57\\
            &           &                    &F814W&4.96e$-$17&19.66& 88& 3.30 x 0.90 (1)&0.73\\
\tableline
\end{tabular}
\end{center}
\small
$^a$ In units of erg s$^{-1}$ cm$^{-2}$ \AA$^{-1}$.\\
$^b$ HST system magnitudes (STMAG).\\
$^c$ Peak surface intensity $I_{s}$ to nebula surface intensity $I_{n}$ ratio.\\
$^d$ Ellipticity $e = 1 - b/a$, where $a$ and $b$ are major- and minor-axis lengths, respectively.\\
$^e$ 2MASS counterpart's coordinates. Only clearly visible part of the nebula was measured on F814W image (see text).
\vspace{0.5cm}
\end{table*}

\begin{table*}[!ht]
\begin{center}
{Table 4: Properties of stellar post-AGB objects}
\small
\begin{tabular}{lrrllrr}
\tableline\tableline
IRAS ID & \multicolumn{2}{c}{Obs. Coord. (J2000)} & \multicolumn{1}{c}{ACS} & \multicolumn{1}{c}{$F_{\lambda}^{~a}$} &
\multicolumn{1}{c}{HST} & \multicolumn{1}{c}{$I_{s}^{~c}$}\\
 & \multicolumn{1}{c}{ra} & \multicolumn{1}{c}{dec} & \multicolumn{1}{c}{filter} & & \multicolumn{1}{c}{Mag$^{b}$}&\\
\tableline
12222$-$4652&12 24 53.49&$-$47 09 08.36&F606W&2.27e$-$12& 8.01~~&1.28e+1  \\
            &           &              &F814W&1.21e$-$12& 8.70~~&4.48e+0  \\
12538$-$2611&12 56 30.00&$-$26 27 36.55&F606W&5.39e$-$12& 7.07~~&1.21e+1  \\
            &           &              &F814W&4.08e$-$12& 7.37~~&1.20e+1  \\
13416$-$6243&13 45 07.26&$-$62 58 15.67&F814W&1.77e$-$15&15.78~~&6.47e$-$6\\
15039$-$4806&15 07 27.52&$-$48 17 53.85&F606W&3.11e$-$12& 7.67~~&1.27e+1  \\
            &           &              &F814W&1.83e$-$12& 8.25~~&6.34e+0  \\
17163$-$3907&17 19 49.26&$-$39 10 38.29&F435W&7.93e$-$16&16.65~~&6.89e$-$3\\
            &           &              &F606W&7.92e$-$14&11.65~~&4.38e$-$1\\
17279$-$1119&17 30 46.82&$-$11 22 08.21&F606W&5.47e$-$13& 9.55~~&3.59e+0  \\
            &           &              &F814W&4.56e$-$13& 9.75~~&5.27e+0  \\
17516$-$2525&17 54 43.35&$-$25 26 28.77&F606W&5.35e$-$16&17.08~~&3.59e$-$3\\
17516$-$2525a$^d$&          &              &     &4.39e$-$16&17.29~~&         \\
17516$-$2525b$^d$&          &              &     &6.12e$-$17&19.43~~&         \\
            &           &              &F814W&9.48e$-$15&13.96~~&3.24e$-$2\\
17516$-$2525a$^d$&          &              &     &7.90e$-$15&14.16~~&         \\
17516$-$2525b$^d$&          &              &     &1.31e$-$16&18.61~~&         \\
17534+2603  &17 55 25.19&  +26 03 01.49&F606W&2.07e$-$11$^e$&5.61$^e$&1.21e+1  \\
            &           &              &F814W&1.12e$-$11$^e$&6.28$^e$&1.23e+1  \\
18135$-$1456&18 16 25.47&$-$14 55 16.17&F606W&5.56e$-$16&17.04~~&4.23e$-$3\\
            &           &              &F814W&6.45e$-$16&16.88~~&3.37e$-$3\\
20547+0247  &20 57 16.29&  +02 58 44.47&F606W&3.85e$-$15$^e$&14.94$^e$&8.32e$-$3\\
            &           &              &F814W&8.20e$-$15&14.12~~&2.35e$-$2\\
\multicolumn{4}{l}{\bf RV Tau stars:}\\
06034+1354  &06 06 15.04&  +13 54 18.95&F606W&7.67e$-$14&11.69~~&4.23e$-$1\\
            &           &              &F814W&1.10e$-$13&11.30~~&3.25e$-$1\\
09256$-$6324&09 26 53.52&$-$63 37 48.78&F606W&1.45e$-$12& 8.49~~&7.80e+0  \\
            &           &              &F814W&1.41e$-$12& 8.53~~&5.27e+0  \\
12067$-$4508&12 09 23.72&$-$45 25 34.60&F606W&6.02e$-$13& 9.45~~&2.89e+0  \\
            &           &              &F814W&4.41e$-$13& 9.79~~&1.93e+0  \\
15469$-$5311&15 50 43.75&$-$53 20 42.88&F606W&2.64e$-$13&10.35~~&1.37e+0  \\
            &           &              &F814W&3.72e$-$13& 9.97~~&1.23e+0  \\
17243$-$4348&17 27 53.60&$-$43 50 46.73&F606W&2.46e$-$13&10.42~~&1.64e+0  \\
            &           &              &F814W&2.55e$-$13&10.38~~&1.07e+0  \\
19125+0343  &19 15 01.17&  +03 48 43.52&F606W&2.71e$-$13&10.32~~&2.39e+0  \\
            &           &              &F814W&2.83e$-$13&10.27~~&1.24e+0  \\
19157$-$0247&19 18 22.75&$-$02 42 10.79&F606W&1.93e$-$13&10.69~~&1.17e+0  \\
            &           &              &F814W&1.68e$-$13&10.84~~&6.11e$-$1\\
20117+1634  &20 14 03.76&  +16 43 35.77&F606W&9.79e$-$13& 8.92~~&5.42e+0  \\
            &           &              &F814W&7.87e$-$13& 9.16~~&3.04e+0  \\
\tableline
\end{tabular}
\end{center}
\small
$^a$ In units of erg s$^{-1}$ cm$^{-2}$ \AA$^{-1}$.\\
$^b$ HST system magnitudes (STMAG).\\
$^c$ Peak intensity (in units of erg s$^{-1}$ cm$^{-2}$ \AA$^{-1}$ sr$^{-1}$).\\
$^d$ See text for details.\\
$^e$ Due to a saturation problem, $F_{\lambda}$ and HST Mag are only the lower limits.
\vspace{0.5cm}
\end{table*}

\subsection{Data reduction and measurements}
Each source was observed with various exposure times in order to increase
the dynamic range of the final image. Faint nebulae were revealed in the
long exposures, but the central star pixels were often saturated. The short
exposures, free from saturated pixels, were combined with the longer ones to
produce a nonsaturated final image. Because each object was observed a 
couple of times we had 2 or more nonsaturated final images. Then IRAF/STSDAS
routines were used for calibration and reduction of images. The MultiDrizzle
package was used to combine multiple frames of images into a single image
(with plate-scale of 0.025$\arcsec$ pixel$^{-1}$), remove cosmic rays and
subtract background radiation.

The reduced images were used to measure photometric and geometric properties
of our objects. We adopted $HST$ photometric calibration of SYNPHOT
\citep[v5.0,][]{lai05} to calibrate the flux density ($F_{\lambda}$ in 
units of erg s$^{-1}$ cm$^{-2}$ \AA$^{-1}$). The defined aperture was big 
enough to encircle the source together with the diffraction features, while 
the background emission was calculated from a (up to, depending on the 
proximity of the other stars on the images) 10-pixel wide annulus separated 
from the selected aperture by a buffer zone. The sky emission was then 
subtracted from the emission of the entire source. The derived quantity, 
which was in counts per second, was finally converted into $F_{\lambda}$ and 
$HST$ ACS magnitudes (STMAG).

We measured the sizes of nebulosities around central stars. Our motivation
was to determine the extension of the nebula as a whole, but not the sizes
of its particular parts (e.g., each lobe). For this we needed only two axis
of extension. In case of DUPLEX sources, where the central star was not
visible and the lobes were clearly separated, we measured the size of each
lobe separately and, in addition, the extension of the whole object. The
extent of the nebula was defined to be the outermost recognizable structure
in which emission level was from 1$\sigma$ up to 14$\sigma$ (depending on
the image quality) above the background level. The major and minor axis, as
well as the ellipticity of the nebulae were derived by fitting ellipses to
the selected isophotes. We measured also surface intensity of the nebulae at
the edge and compared it to the star (or peak if the central star was not
visible, as in DUPLEX sources) surface intensity and got star-to-nebula
surface intensity ratios ($I_s / I_n$). For objects without nebulae only the
peak intensity was measured. Names and the coordinates of all objects, as
well as other derived parameters ($F_{\lambda}$, $HST$ magnitudes, $I_s /
I_n$ and extension) are shown in Tables 2, 3 and 4 for SOLE, DUPLEX and
stellar sources, respectively.

Figures 1, 2 and 3 show the reduced images. The logarithmic scale is used to
illustrate the contrast between nebulae and central stars. Nevertheless,
some faint nebulae are still barely visible. $I_s / I_n$ ranges from 4 to $3
\times 10^5$ (see Tables 2 and 3). The point spread function (PSF) effects
are seen on most of the images. To remove them and to make the nebulae more
clearly seen we deconvolved our images using Richardson-Lucy (RL) algorithm
with a stellar PSF created especially for our images by the code TinyTim
\citep[v6.3,][]{kri04}. As a result, the contrast between nebula and the
background emission was improved in general, but in case of very faint
nebulosities where the nebulae emission was comparable to sky emission, the
difference between reduced and deconvolved images is rather insignificant.
There is also one inconvenience in using RL algorithm, which is known to
produce false depressions around unusually bright pixels. Those ``holes''
are seen on our deconvolved images, but are not considered real structures.
The deconvolved images are shown together with the reduced ones in Figures
1, 2 and 3 in the same logarithmic scale.

\begin{figure*}[!ht]
\includegraphics[width=\textwidth]{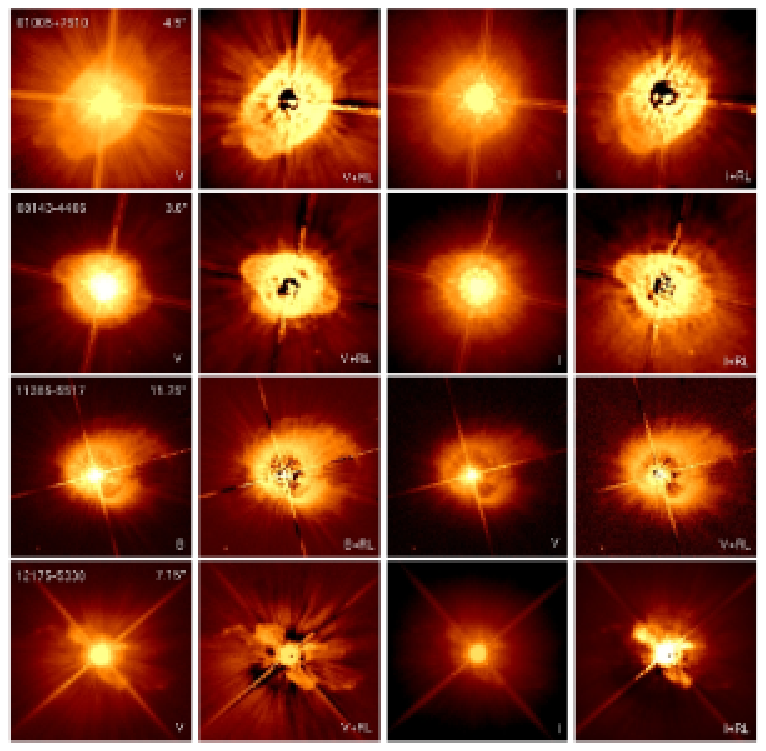}
{Fig.1. - Images of SOLE objects (north is up and east is to the left)
shown in order of increasing RA (see Table 2). The leftmost frame shows the
IRAS ID and size of an image. The filters types are shown at the bottom of
each frame with ``+RL'' indicating Richardson-Lucy deconvolution. The images
are in logarithmic scale to illustrate the contrast between central star and
the nebula.} \label{imsole1}
\vspace{0.5cm}
\end{figure*}

\begin{figure*}[!ht]
\includegraphics[width=\textwidth]{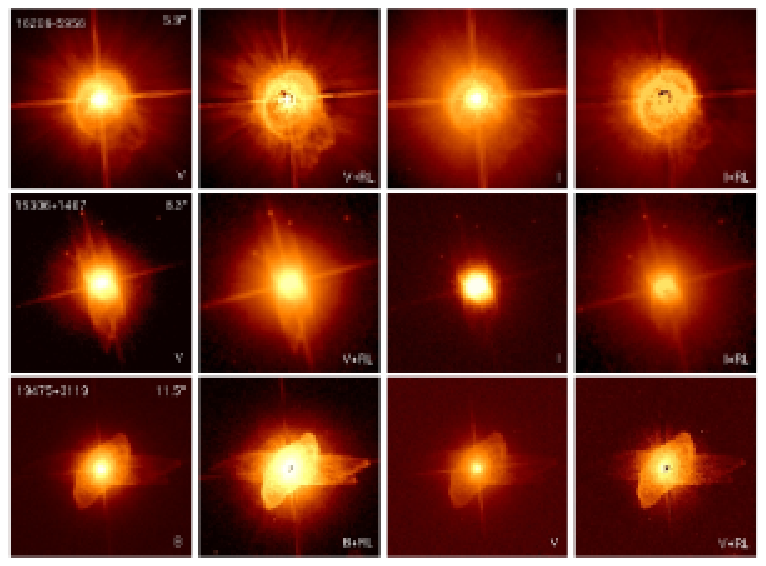}
{Fig.1. - continued.}
\label{imsole2}
\vspace{0.5cm}
\end{figure*}

\begin{figure*}[!ht]
\includegraphics[width=\textwidth]{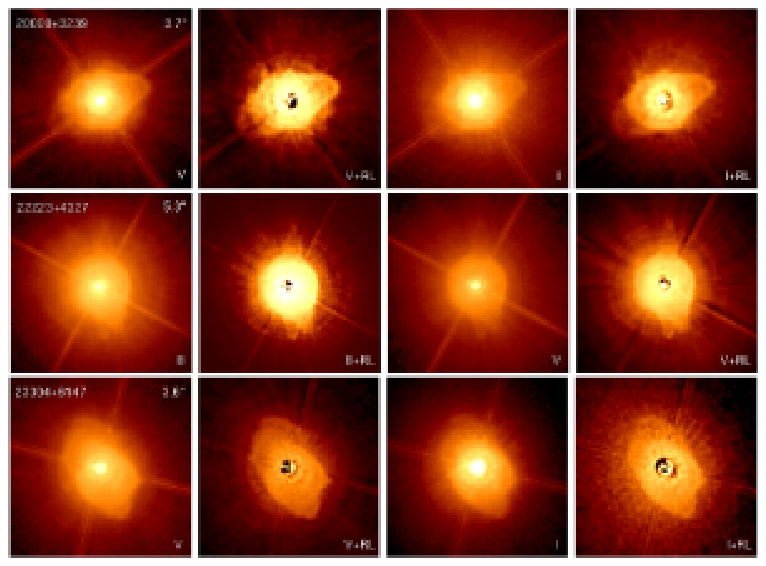}
{Fig.1. - continued.}
\label{imsole3}
\vspace{0.5cm}
\end{figure*}

\begin{figure*}[!ht]
\includegraphics[width=\textwidth]{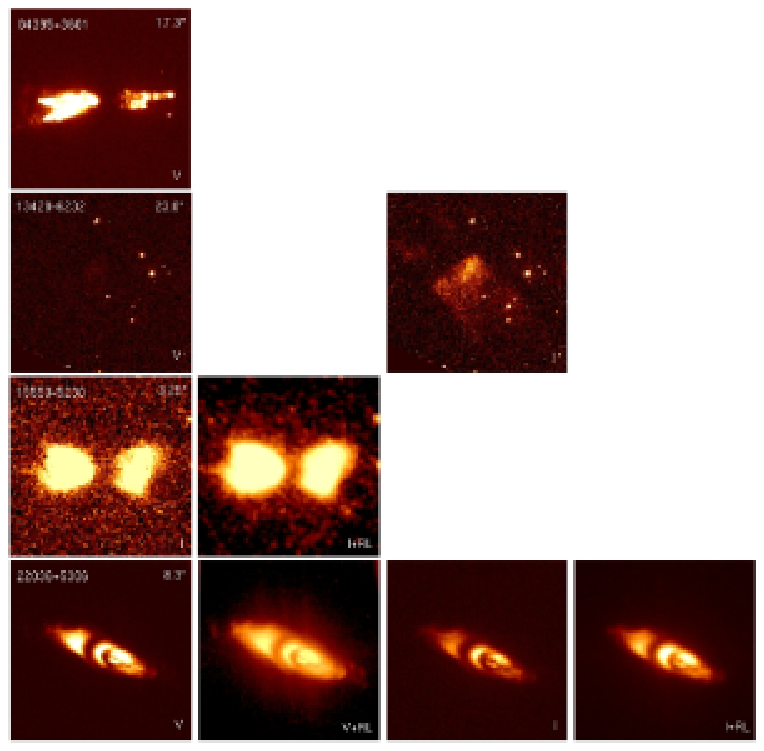}
\label{imduplex1}
\end{figure*}

\begin{figure*}[!ht]
\vspace{-0.7cm}
\includegraphics[width=\textwidth]{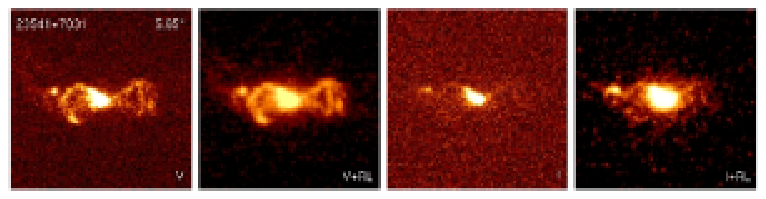}
{Fig.2. - Images of DUPLEX objects shown in order of increasing RA (see
Table 3). The displaying scheme is the same as in Fig.1.}
\label{imduplex2}
\vspace{0.5cm}
\end{figure*}

\section{Results}
We detected nebulosities in 15 out of 33 sources and classified them as SOLE
and DUPLEX following UMB00. SOLEs have an optically thin (continuous)
envelope and the star is therefore very bright at optical and near-infrared
wavelengths in comparison to the nebula (in terms of $I_s / I_n$ we consider
values of $>1000$). The surface brightness peaks at the central star.
Because of multiaxis lobes, with different shapes and surface brightness,
UMB00 divided those objects in subgroups of simple ellipse, multiple
ellipses, an ellipse with embedded bipolar structure and an ellipse with
concentric shells. These structures are also present in our objects. In
addition we observe snail-like structures. The PSF effects are apparent in
the images: the faint nebular structures are often buried under the PSF of
the bright central star. DUPLEXes have an optically thick torus that
completely or partially obscures the central star. $I_s / I_n$ values are
much below 1000 and the peak of the surface brightness is far away from the
center of the object. The nebula is bipolar and the (multiple) lobes are
seen through scattered light. The PSF diffraction feature is usually not
seen in the images of DUPLEX objects.

These observational characteristics of the nebulae are consistent with the
present understanding of the structure of these shells, in which SOLE and
DUPLEX nebulae respectively represent the low and high ends of the degree of
equatorial enhancement for an intrinsically axisymmetric circumstellar
shell. In other words, there is not much of a difference between SOLE and
DUPLEX nebulae as far as the shell structure goes, but the difference is the
optical depth along the equator due to the difference in the equator-to-pole
density ratio resulting from the different degree of equatorial density
enhancement. This is what makes DUPLEX nebulae appear bipolar with a
well-defined dust lane, while making SOLE nebulae appear elliptical without
a clear dust lane. Because of this, their shell morphology tends to become
less definitive when the optical depth along the equator is close to unity
(and/or the inclination angle of the shell is far from edge-on).

We have 10 SOLE objects (IRAS 01005+7910, IRAS 08143$-$4406, IRAS
11385$-$5517, IRAS 12175$-$5338, IRAS 16206$-$5956, IRAS 19306+1407, IRAS
19475+3119, IRAS 20000+3239, IRAS 22223+4327 and IRAS 23304+6147) and 5
DUPLEX objects (IRAS 04395+3601, IRAS 13428$-$6232, IRAS 15553$-$5230, IRAS
22036+5306 and IRAS 23541+7031) in our sample. $I_s / I_n$ for our DUPLEXes
is at most 330 (IRAS 22036+5306) and for SOLEs reaches the value of a few
hundred thousands (e.g., IRAS 11385$-$5517 or IRAS 19475+3119). The
remaining 18 sources do not show any nebulosities around them and are
referred to as stellar objects. Due to a different morphology we may expect
also the differences in spectral energy distribution (SED) shapes and the
position in infrared color-color diagrams. All groups of sources are
discussed separately in the following sections. Images of SOLE, DUPLEX and
stellar objects are shown in Figures 1a, 1b and 1c, respectively.

\subsection{SOLE objects}
IRAS 01005+7910 have a nebulosity of the similar size in both bands
(3.75$\arcsec$ $\times$ 2.20$\arcsec$ in $V$ and 3.65$\arcsec$ $\times$
2.25$\arcsec$ in $I$). Its morphology is irregular. It has two lobes of
different shapes pointed in north-west and south-east and additional small
lobe emerging in south-eastern direction visible in the $V$ band. The
circular halo corresponds to the wings of the PSF around the bright central
star.

IRAS 08143$-$4406 shows an elliptical nebula of 2.1$\arcsec$ long and
1.4$\arcsec$ wide. It has two small lobes emerging in western and eastern
direction. The circular PSF halo is also present. The object looks similar
in both bands with the only exception in $I_s / I_n$, which is two times
bigger in the $V$ band than in the $I$ band. This source is a carbon-rich
object with the overabundance of s-process elements \citep{rey04} and is
similar in appearance to IRAS 07430+1115 from UMB00 sample, which is also
rich in s-process elements \citep[e.g][]{red99}.

The nebulosity of IRAS 11385$-$5517 is very asymmetric and looks like
pinwheel. This A-type oxygen-rich low-mass supergiant was thought to be a
binary, but later observations suggested rather disk-like structure and
high-velocity bipolar outflows \citep[e.g.,][]{olo99}. It is possible that
we see only one of this outflows on $HST$ images because of the inclination
effect. The snail-like shape of the outflow is quite similar to the spiral
pattern found in C-rich AGB star AFGL 3068 \citep{mau06}. The spiral 
outflow in IRAS 11385$-$5517 is also single-armed and its extension is 
8.3$\arcsec$ $\times$ 6.6$\arcsec$ in the $B$ band. The object is very 
bright ($B$=6.6mag and $V$=7.3mag) and looks similar in both bands. $I_s / 
I_n$ is very high in the $B$ band with the value of $\sim 3.4 \times 10^5$ 
being the highest among SOLE sources.

IRAS 12175$-$5338 has very complex morphology. The nebula is 5.7$\arcsec$
long and 2.5$\arcsec$ wide. It has at least 4 lobes, pointed in different
directions. The lobes are of very different brightness, with the bright ones
closer to the central star and very faint more extended structures. They are
less clearly visible in the $I$ band. The circular PSF halo is also seen
around the star. The object is quite bright in both bands, $\sim$9.5mag, and
is known to be a metal-deficient pulsating variable \citep[e.g.,][]{win97}.

IRAS 16206$-$5956 has a nebulosity very similar to the one seen in IRAS
11385$-$5517. It is also asymmetric and has a snail-like shape, but is much
smaller with an extension of 3.85$\arcsec$ $\times$ 2.6$\arcsec$. There is
also a faint lobe emerging to the south-west with a quite complex structure.
There is no visible lobe on the other side of the objects, but it may be
extremely faint and buried under the PSF of the bright central star. The
structures are better resolved in the $V$ band than in the $I$ band, but
$I_s / I_n$ is the same, $\sim$21000, in both bands. IRAS 16206$-$5956 is
also A-type oxygen-rich supergiants similar to IRAS 11385$-$5517, but is
more massive with a core mass of $M=0.66M_{\odot}$ (see Table 1).

IRAS 19306+1407 is a B-type oxygen-rich PPN. It has a bright central star
and low surface brightness lobes ($I_s / I_n$ is $\sim$ 9000 and 6400 in the
$V$ and $I$ band, respectively). The lobes have bipolar structure, also seen
in previous observations \citep{volk04,sah04}. The lobes seem to be smaller
in the $I$ band, with the extension of only 4.4$\arcsec$ $\times$
2.0$\arcsec$, in comparison to the $V$ band where they are longer,
7.7$\arcsec$ $\times$ 2.0$\arcsec$.

IRAS 19475+3119 has a quadrupolar nebula and a PSF spherical halo. Similar
complex double-elongation structure can be seen in IRAS 04296+3429 from
UMB00, except for the different sizes of the nebula: IRAS 19475+3119 has
much bigger nebulosity of 10.45$\arcsec$ long and 5.1$\arcsec$ wide. Also,
IRAS 19475+3119 is known to be oxygen-rich \citep[its dust envelope was
described in details by][]{sar06}, whereas IRAS 04296+3429 is a carbon-rich
source with s-process overabundance and 21 $\mu$m emission feature
\citep[e.g.,][]{volk99}.

IRAS 20000+3239 is a carbon-rich source with s-process elements and 21 
$\mu$m feature \citep[e.g.,][]{volk99,win00a}. It has small lobes emerging
to west and east with the extension of 2.15$\arcsec$ $\times$ 1.45$\arcsec$
in $V$ (and slightly smaller in $I$ band) and is very similar to IRAS
08143$-$4406 described above.

IRAS 22223+4327 is also a carbon-rich source with s-process elements and 21
$\mu$m feature. It is quite bright in $V$ band ($\sim$9.6) and the nebula
has a size of 3.4$\arcsec$ x 2.1$\arcsec$. There are two pairs of lobes
pointed to north and south and they are more clearly visible on the south
side. The object is quite similar to IRAS 08143$-$4406 and IRAS 20000+3239.

IRAS 23304+6147 also belongs to the group of objects with 21 $\mu$m feature.
Its extension is comparable to the extension of IRAS 20000+3239
(2.25$\arcsec$ $\times$ 1.6$\arcsec$). It has 2 lobes emerging in northern
and southern directions and another inner pair of lobes in western and 
eastern directions. Those inner lobes are not so clearly seen even on
deconvolved images. The object's structure could be similar to one seen in
IRAS 06530$-$0213 from UMB00. While the appearance of IRAS 20000+3239,
IRAS 22223+4327 and IRAS 23304+6147 is similar and those three objects have
the same 21 $\mu$m feature, the SOLE morphology is not characteristic to the
whole group with 21 $\mu$m, e.g., DUPLEX source IRAS 22574+6609 (UMB00) or
Egg Nebula \citep[e.g.,][]{sah98}.

\subsection{DUPLEX objects}
IRAS 04395+3601 (better known as AFGL 618) is a carbon-rich PPN surrounding
a B0 central star, rapidly evolving to a PN stage. The ACS image shows
bipolar structure of the nebula, with the extent of 15.4$\arcsec$ $\times$
4.8$\arcsec$. The eastern lobe is larger (8.25$\arcsec$ $\times$
4.75$\arcsec$) and brighter ($V$=15.7mag) than western one (6.2$\arcsec$
$\times$ 3.2$\arcsec$ and $V$=17.1mag). The star itself is hidden in the
thick dust torus and only huge outflows are seen in the optical
\citep{tram02}. The structure of the object is very complex. The lobes
consist of shock-excited gas and dust and are composed of several jetlike
structures \citep[e.g.,][]{san02}. The compact HII region is present in the
central part of the nebula and the circumstellar envelope contains various
molecular species \citep[e.g.,][]{woo03a,woo03b}.

IRAS 13428$-$6232 is a very faint object with no optical counterpart. The
nebula is not visible in $V$ band and parts of it are revealed in $I$ band.
The object is extended in the north-east and south-west directions with a
size of more than 20$\arcsec$. However, only its brightest central fragment
with the size of 8.75$\arcsec$ $\times$ 6.9$\arcsec$ (extended
perpendicularly to the whole object) is above 1$\sigma$ level and could be
measured. This source was studied previously by \citet{vds00}, who observed
a nice bipolar shape of IRAS 13428$-$6232 in the near-infrared with the
angular size of 4.1$\arcsec$ $\times$ 12.0$\arcsec$ in $N$ band and
6.0$\arcsec$ $\times$ 11.0$\arcsec$ in $K$ band.

IRAS 15553$-$5230 is a poorly known post-AGB star. This IRAS source was
observed in near-infrared by \citet{gar97} and \citet{vds00} and there was
no agreement about the correct counterpart, however, \citet{vds00} noticed
that the object was elongated. ISO spectrum shows no features. Our
observations confirmed coordinates indicated by \citet{vds00} and revealed
the bipolar shape of this small (2.5$\arcsec$ x 1.1$\arcsec$) nebula. The
object is very faint and the central star is invisible. The lobes differ in
size and shape, which may be the effect of the nebula orientation towards
observer. A small feature pointing out from the east lobe may be a faint
outflow or a jet.

\begin{figure}[!ht]
\includegraphics[width=0.478\textwidth]{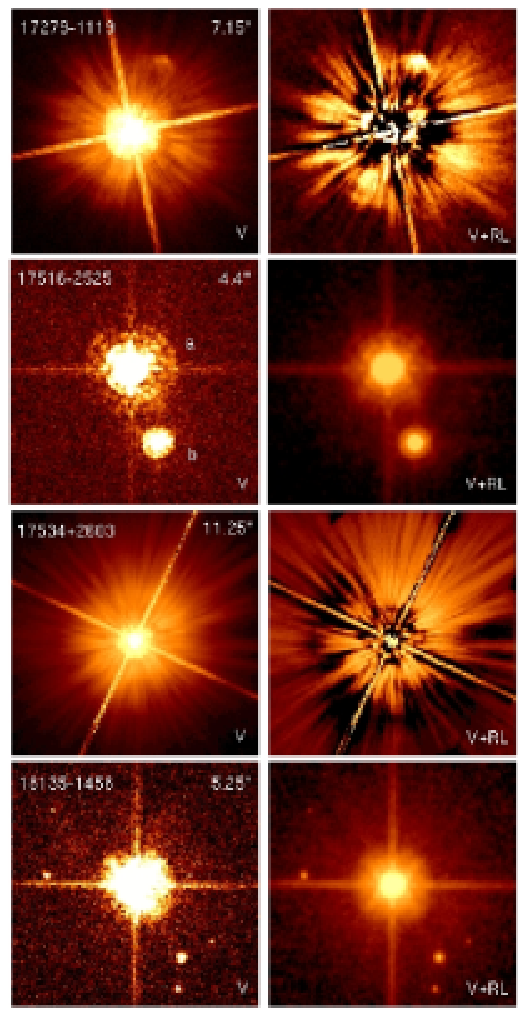}
{Fig.3. - Exemplary images (reduced and deconvolved) of 4 stellar objects.
IRAS name, filter and size of images are shown.} \label{imstellar}
\vspace{0.5cm}
\end{figure}

The extended and complex bipolar nebula of IRAS 22036+5306 was already known
from the previous $HST$ observations \citep[proposal ID 9101,][]{sah03}. 
The size of the object and magnitudes in the $V$ and $I$ bands determined in
this paper are slightly different from the ones given by Sahai and
collaborators, but the WFPC2 camera used in their observations had different
plate scale and different response curve than ACS. The morphology of the
object is very complex, with central ring structures, elongated lobes and
faint knotty jetlike linear structures. The object is oxygen-rich, with a
very characteristic feature in ISO spectra of water ice absorption at $\sim$
3.1 $\mu$m \citep[e.g.,][]{sah03}.

IRAS 23541+7031 (also known as M 2$-$56) is a PPN with a strong shock
emission seen in optical line spectrum. \citet{tram98} discovered its
bipolar nature and found weak emission extending up to 10$\arcsec$ from the
central bright region. From our measurements the nebula is only
3.75$\arcsec$ x 1.6$\arcsec$ in size in the $V$ band and smaller and much
fainter in the $I$ band. However, there is a very faint emission (below
1$\sigma$ above the sky level) outside the measured size and we were not
able to estimate its extent. The central star of IRAS 23541+7031, which is
much fainter in the $I$ band is classified as type B, but cooler than the
central star of AFGL 618.

\subsection{Stellar objects}
Stellar sources are not associated with any obvious nebulosities. Objects
are both very bright, like IRAS 17534+2603, and very faint, like IRAS
18135$-$1456. For bright sources diffraction features are very strong.
Various artificial features produced during deconvolution can be seen in
Figure~3.

All analyzed RV Tau stars belong to this group. They are cool post-AGB
objects with effective temperatures below $8000K$, usually low-mass central
stars and oxygen-rich envelopes \citep[for few exceptions see
e.g.,][]{sta06} with photospheres depleted in the high condensation
temperature elements as a result of gas-dust separation
\citep[e.g.,][]{gon97}. Their IR excess indicates the presence of a warm
dust which should be visible as an extended circumstellar envelope. However, 
there is no evidence of such envelope on analyzed images. This is caused by 
properties of both $HST$ and RV Tau stars. $HST$ is sensitive to only 
nearby, relatively dense circumstellar dust, while RV Tau stars (because of 
their long-time evolution) have very extended and therefore very low surface 
brightness envelopes. Hence, RV Tau stars appear as stellar sources.

The rest of the objects in the stellar group are post-AGB stars with
infrared excess. IRAS 12538$-$2611 and IRAS 15039$-$4806 are supergiants
with cool dust, while IRAS 12222$-$4652 and IRAS 17279$-$1119 show a near-IR
excess which stands for hot dust (compare SEDs on Fig.6.). It was suggested
by \citet{win00} that all objects with near-infrared excess are binaries.
One of them is also IRAS 17534+2603, better known as 89 Her. It's a
high-latitude F2 supergiant with a hot dust in a shell, large CO shell
\citep{fong06} and the H$\alpha$ line showing the evidence for the on-going
mass loss \citep[e.g.,][]{wat93}. The model constructed for the
circumstellar envelope also assumed that IRAS 17534+2603 must have evolved
as a binary system. Only for IRAS 17279$-$1119 one can see a clump of denser
material in the north-west part of an image. It must be connected with the
star since there are no other objects in the near vicinity. It may be an
ejected knot. The object is classified as stellar source since we do not see
any evidence for continuous nebulosity around it (as it is in case of SOLE
objects), but the knot makes it peculiar.

IRAS 13416$-$6243 was connected with a highly reddened G1 supergiant by
\citet{hu93}, but the conclusions from low resolution IRAS spectrum about 
10 and 18 $\mu$m silicate absorption bands were misleading. Instead, ISO
spectrum shows moderate PAH bands and absorption feature of C$_{2}$H$_{2}$
at 13.7$\mu$m suggesting a carbon rich, dusty envelope. There is also a
6.9$\mu$m feature which is seen only in the post-AGB phase and only in
objects with a 21$\mu$m feature \citep[e.g.,][]{hriv00}. Unfortunately, a
21$\mu$m is not seen in this object and we can only speculate that the
feature is quite weak and ISO resolution was not enough to see it.

IRAS 17516$-$2525 is known to be an oxygen-rich source with the OH maser
emission. It is a very faint star with $V \sim$17.8mag, but analyzed images
show that there are indeed 2 stars very close to each other! The separation
is 1.45$\arcsec$ and they are resolved only on $HST$ images. It is not clear
if the stars are related. IRAS 17516$-$2525a is redder in the $I$ band than
IRAS 17516$-$2525b and most likely this is a correct IRAS counterpart.

IRAS 20547+0247 is a peculiar variable star. IRAS 12/25 color suggests the
optically thin envelope, however IRAS-LRS spectrum shows 10$\mu$m silicate
absorption feature typical for thick dusty envelope. To explain this
contradiction \citet{bar96} proposed a dusty, thick disk around IRAS
20547+0247 but seen edge-on. Its optical spectrum is very unusual with
strong absorption and emission bands of metallic oxides, TiO, AlO, and VO,
which should be not visible in G-type star. The infrared spectroscopy
revealed the hot circumstellar gas located probably in a disk-like structure
\citep{geb05}. We see no evidence for a disk in the images.

\section{Discussion}
In this section we describe various properties of the analyzed objects.
Table 5 compares the morphology of the objects with other properties. To get
a wider view we include in our discussion other post-AGB objects observed
with $HST$ and already analyzed by other authors. Together there are 66
post-AGB objects with $HST$ images published in the literature.

\subsection{Nebulosities}
The morphologies of PPNs are very complex for both SOLE and DUPLEX objects,
from an optically thin nebulosities to multiple dusty outflows with
microstructures as we have reviewed in $\S$3. They are different for SOLE
and DUPLEX objects, but they all show some degree of asymmetry. All nebulae
are elongated. The sizes of the nebulae vary from small (e.g., 2.1$\arcsec$
$\times$ 1.3$\arcsec$ for IRAS 08143$-$4406) to big ones (e.g., IRAS
04395+3601 with the extent of 15.4$\arcsec$ $\times$ 4.8$\arcsec$), but the
physical extension is not known due to a lack of known distances to the
objects. Small or faint nebulosities can be also buried under PSF of bright
central star and therefore hard to see. The surface intensity ratio between
star and nebula in SOLE objects is always high, varying from 2000 to almost
340000 and the nebula appears faint in comparison to the central star.
DUPLEX sources are all faint ($V$=15--21mag in this study) and the surface
intensity ratio is below 330. Hence, even huge nebulae like in AFGL 618 are
not easy to detect.

As mentioned already in this paper, the observational characteristics of the
nebulae are consistent with the present understanding of the structure of
these shells, in which SOLE and DUPLEX nebulae respectively represent the
low and high ends of the degree of equatorial enhancement for an
intrinsically axisymmetric circumstellar shell. As far as the shell
structure goes, there is not much of a difference between SOLE and DUPLEX
nebulae, but the difference is the optical depth along the equator due to
the difference in the equator-to-pole density ratio resulting from the
different degree of equatorial density enhancement. Thus DUPLEX nebulae
appear bipolar with a well-defined dust lane, while SOLE nebulae appear
elliptical without a clear dust lane. However, their shell morphology tends
to become less definitive when the optical depth along the equator is close
to unity and/or the inclination angle of the shell is far from edge-on.

Table 5 shows that there is no obvious connection between morphology and
characteristics of the objects in our sample. Among SOLEs we have a
comparable number of carbon- and oxygen-rich objects which implies that the
shapes of the nebulae are not necessary correlated with the chemical
composition of the objects. Good examples are two oxygen-rich supergiants 
IRAS 11385$-$5517 and IRAS 16206$-$5956 and a carbon-rich AGB star AFGL 3068 
(see $\S$3.1). 

\begin{figure*}[!ht]
\includegraphics[width=\textwidth]{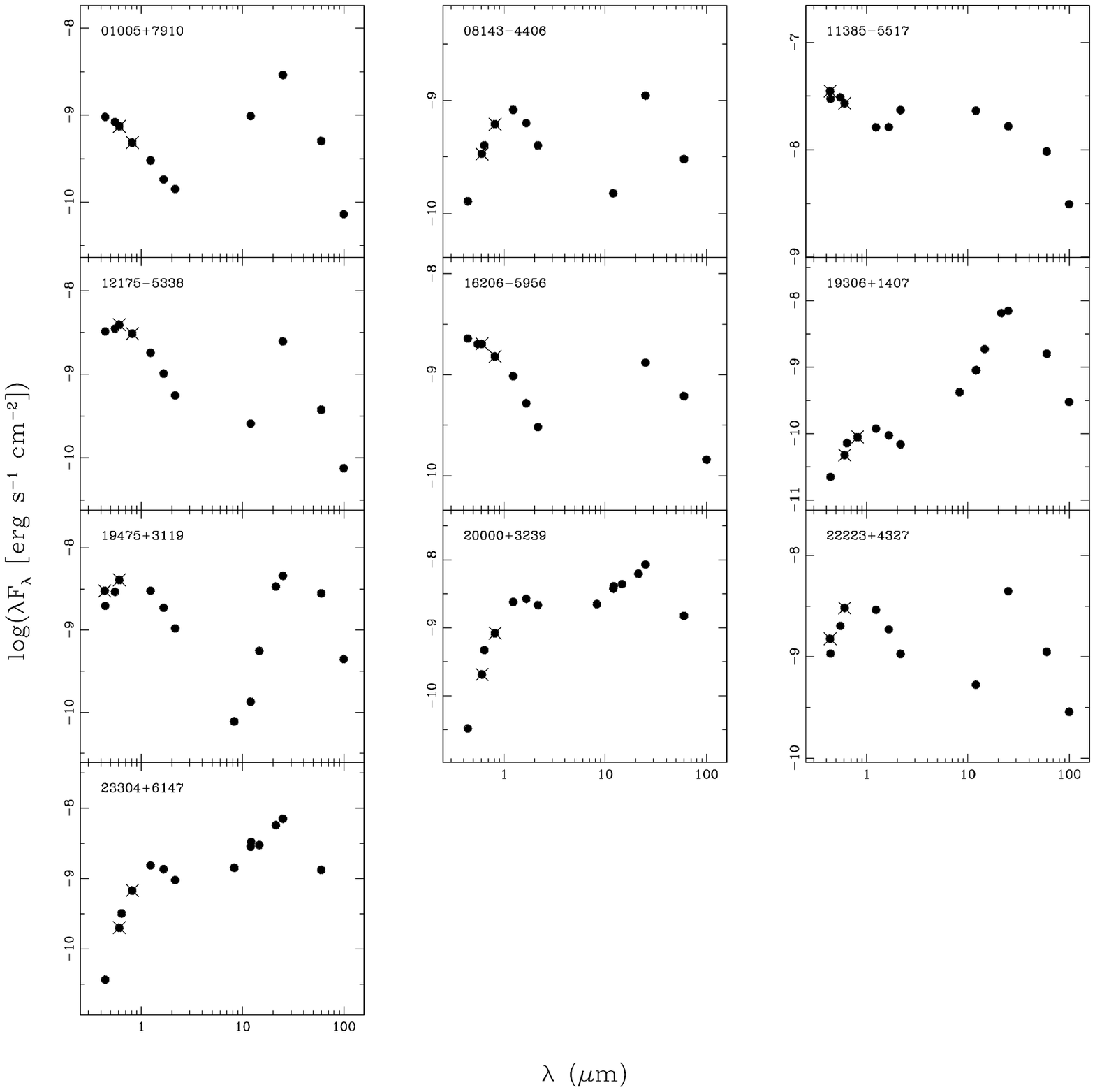}
{Fig.4. - Spectral energy distributions for SOLE objects shown in order of
increasing RA. Values are taken from GSC2.2, 2MASS, MSX6C and IRAS
catalogues. Measured HST magnitudes are shown as crossed points.}
\label{sed1}
\vspace{0.5cm}
\end{figure*}

\begin{figure*}[!ht]
\includegraphics[width=\textwidth]{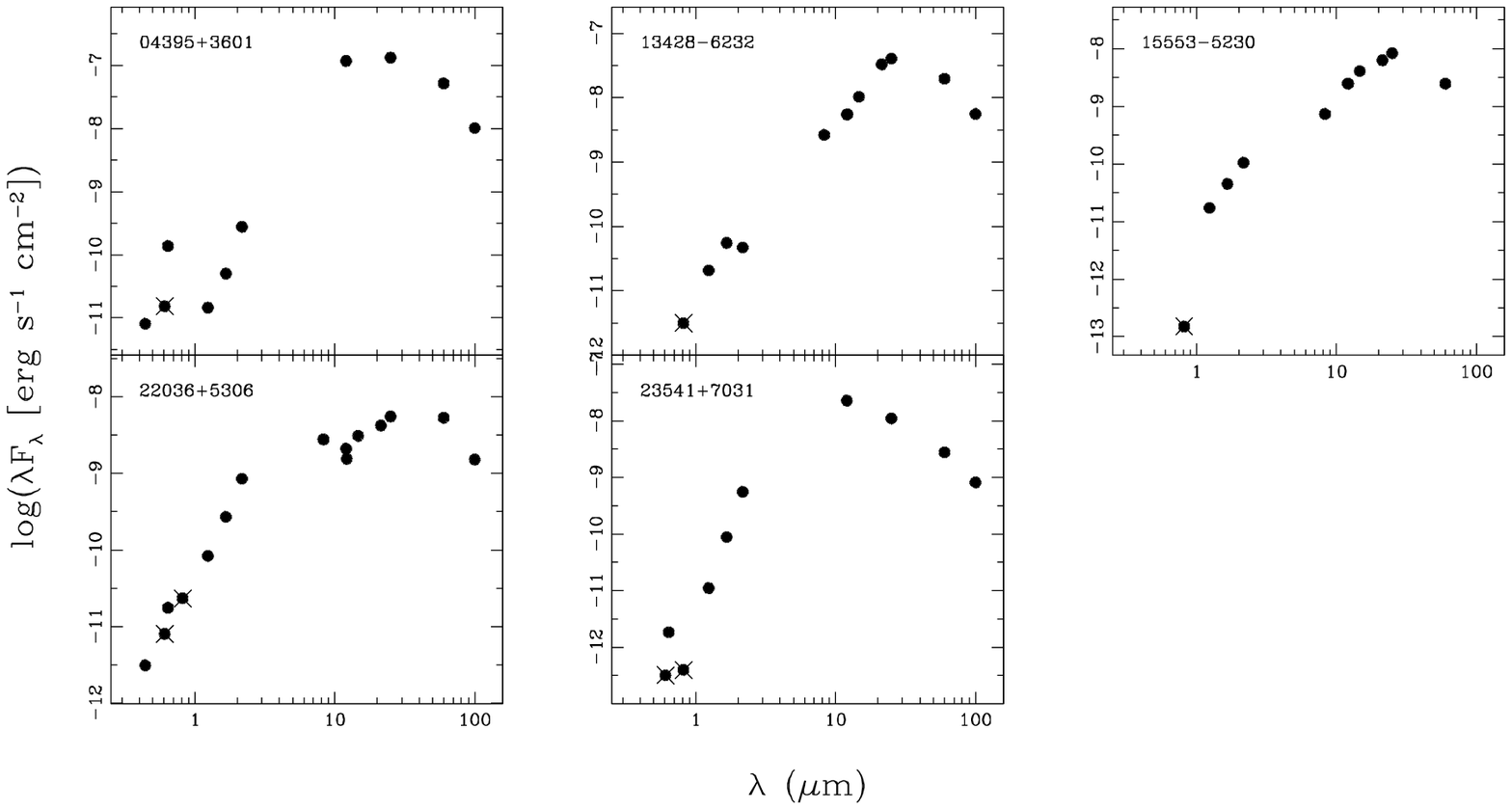}
{Fig.5. - Spectral energy distributions for DUPLEX objects shown in order of
increasing RA. The displaying scheme is the same as in Figure 4.}
\label{sed2}
\vspace{0.5cm}
\end{figure*}

\begin{figure*}[!ht]
\includegraphics[width=\textwidth]{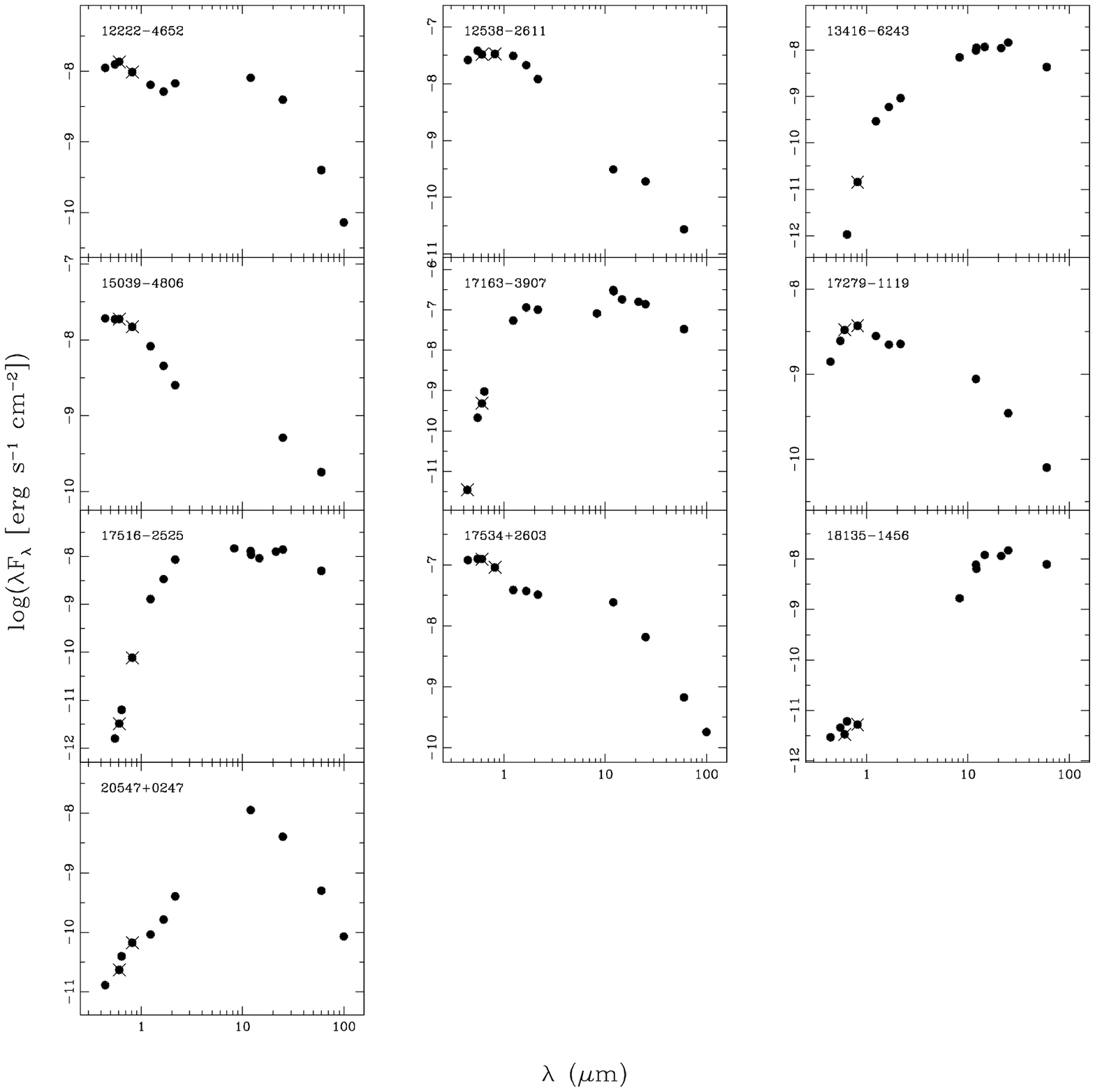}
{Fig.6. - Spectral energy distributions for stellar sources shown in order
of increasing RA. The displaying scheme is the same as in Figure 4. SED of
IRAS 18135$-$1456 is composed of two sources, see explanation in text, $\S$4.3.}
\label{sed3}
\vspace{0.5cm}
\end{figure*}

\begin{figure*}[!ht]
\includegraphics[width=\textwidth]{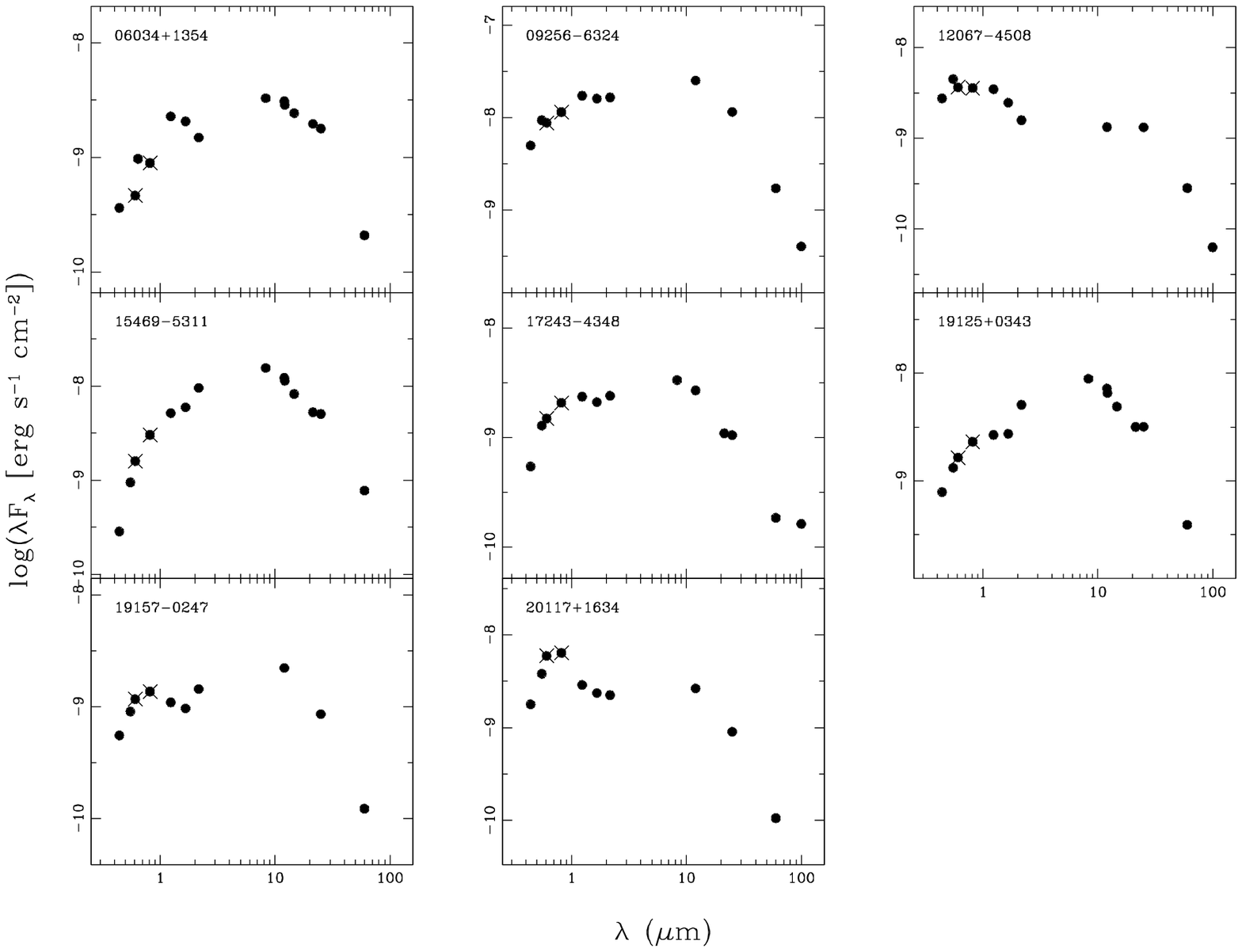}
{Fig.7. - Spectral energy distributions for RV Tau objects shown in order of
increasing RA. The displaying scheme is the same as in Figure 4.}
\label{sed4}
\vspace{0.5cm}
\end{figure*}

Also, SOLE objects occupy a wide range of effective temperatures
($5000K<T_{\rm eff}<25000K$) and masses ($0.55M_{\odot}<M<0.91M_{\odot}$).
Therefore, it is unlikely that SOLE morphology represents a specific
evolutionary epoch and a specific mass object. The lack of known chemical
composition and masses of most DUPLEXes makes it impossible to connect the
morphology with these properties. The range of $T_{\rm eff}$ of DUPLEXes is 
large (K7 - B, see Table 5) and no correlation with morphology can be found. 
On the other hand, we find both morphologies among sources with similar 
chemical and physical properties, e.g. 21 $\mu$m sources: IRAS 20000+3239
and IRAS 22223+4327 are SOLEs and Egg Nebula and IRAS 06176$-$1036 are
DUPLEXes. UMB00 also didn't find any connections between shapes of the
nebulae and properties of analyzed objects.

\subsection{Magnitudes of analyzed objects}
Derived $B$, $V$, and $I$ $HST$ magnitudes are in general consistent with
previously published measurements, although ACS filters do not have exact
counterparts in other photometric systems \citep[photometric 
transformations were discussed by][]{sir05}. However, 2 DUPLEX objects, IRAS 
04395+3601 and IRAS 23541+7031, are very bright in the $R$ band in 
comparison to other optical wavelengths (compare also SEDs on Fig.5). AFGL 
618 was not measured previously in the $V$ band. Our measurements show it is 
faint at $V_{ACS}$=15.39. From SED (Fig.5) we see that it is also quite 
faint in the near-infrared. However, the object has significant line 
emission in H$\alpha$ \citep[e.g.,][]{san02} and that may account for the 
anomalous increase in the $R$-band brightness. Similarly, IRAS 23541+7031 is 
a bright source in H$\alpha$ \citep[e.g.,][]{tram98} and its $R$ magnitude 
is higher than expected. The H$\alpha$ line is an evidence for the ongoing 
mass loss and can be found in many post-AGB objects, even in RV Tau stars
\citep[e.g.,][]{maas05}. We do not see such a big increase of brightness in
the $R$ band in any other object in our sample.

\subsection{Spectral energy distribution}
Due to differences between SOLEs and DUPLEXes we expect them also to have
different SED shapes, as was already discussed by UMB00. While SOLEs show
both of the optical and IR peaks clearly, DUPLEXes show the prominent IR
peak with a highly obscured optical peak. This is consistent with DUPLEXes
being optically thicker than SOLEs due to their density distribution of a
higher equator-to-pole ratio. SEDs for 33 objects which images are analyzed
in this study are shown in Figures 4, 5, 6 and 7. Photometric values are
taken from GSC2.2, 2MASS, MSX6C, IRAS catalogues and HST measurements. We
adopted the classification scheme of \citet{veen89} and grouped objects into
four groups: I - flat spectrum between 4 and 25 $\mu$m and a steep fall-off
toward shorter wavelengths, II - maximum around 25 $\mu$m and a gradual
fall-off to shorter wavelengths, III - maximum around 25 $\mu$m and a steep
fall-off to shorter wavelengths with a plateau between 1 and 4 $\mu$m, IV -
two distinct maxima: one around 25 $\mu$m and second between 1 and 2 $\mu$m
(IVa) or below 1 $\mu$m (IVb).

In SOLE objects, the optical peak due to the central star and the infrared
peak due to dust emission are clearly distinguishable. IRAS 08143$-$4406,
IRAS 20000+3239, and IRAS 23304+6147 have the optical maximum above 1 
$\mu$m and their nebulae are small (compare the sizes in Table 2) and they 
are of class IVa. The rest of SOLE sources, except for IRAS 19306+1407, are 
of class IVb with a peak below 1 $\mu$m and they have extended nebulosities. 
In addition, SED of IRAS 11385$-$5517 shows a prominent near-infrared excess
that suggests a hot dust in circumstellar material and/or ongoing mass loss.
SED diagram for IRAS 19306+1407 is slightly different from other SOLE
objects, with a very strong cold dust emission in comparison to the stellar
component, but no indication of a hot dust which could be stored in the
vicinity of the central star. The size of the nebula is rather big, which is
characteristic for class IVb, but the optical maximum seems to be above 1
$\mu$m like a class IVa object. It may be a case of an intermediate optical
depth object with a massive circumstellar envelope that makes its dust peak
higher than the stellar component. The lack of other objects with
intermediate optical depth in our sample is caused by the fast evolution of
post-AGB objects with effective temperature between 10000 and 16000
\citep[e.g.,][]{sch93} and hence a very small number of those objects to
observe.

SEDs for DUPLEX objects are different from those for SOLE sources. The dust
maximum is very prominent and the optical part is steep or flat without a
clear peak. SEDs are of class II (for IRAS 22036+5306 and IRAS 23541+7031)
and III (for IRAS 04395+3601 and IRAS 13428$-$6232) depending on the
steepness of fall-off at shorter wavelengths. We have classified IRAS
15553$-$5230 as type II/III.

Stellar sources show various kinds of SEDs. The non RV Tau sources can be
divided into two groups. IRAS 12222$-$4652, IRAS 12538$-$2611, IRAS
17279$-$1119, and IRAS 17534+2603 have relatively flat optical/near-infrared
spectrum decreasing in the far-infrared. One may see some resemblance of 
their SEDs to the SED of SOLE object IRAS 11385$-$5517. This indicates that 
these sources are actually SOLE objects without nebulosities, and hence, 
viewed pole-on and/or too distant and/or too compact for the shell to be 
resolved. IRAS 15039$-$4806 may also belong to this category, but its SED is 
much steeper in the mid-infrared. The second group consists of IRAS 
13416$-$6243, IRAS 17163$-$3907, IRAS 17516$-$2525, and IRAS 20547+0247 with 
a steep, rising near/mid-infrared spectrum and prominent but flat 
far-infrared part. They look similar to SEDs of DUPLEXes, which may suggest 
that they are indeed DUPLEX sources. However, we do not see any apparent 
presence of circumstellar envelope/disk on $HST$ images. The lack of visible
nebulosities may be caused by an orientation effect as the sources with the
possible tori around are viewed almost pole-on. Dust scattering is typically 
anisotropic and directions of scattering almost perpendicular to the
direction of incident photons are less probable \citep[e.g.,][]{dra03}.
Therefore, pole-on DUPLEX sources may not show obvious nebulosities.
Those objects may also be too compact and/or too distant to be resolved and
hence appearing as stellar sources. In fact, IRAS 13416$-$6243 is known to
have a dusty envelope and IRAS 20547+0247 is suspected to have dusty thick
disk around central star seen edge-on (see previous sections). SED of IRAS
17516$-$2525 is similar to the ones of DUPLEX objects, but it is constructed
from the combined magnitudes of two stars unresolved up to now and the
contribution from each star cannot be separated.

\begin{figure*}[ht]
\includegraphics[width=\textwidth]{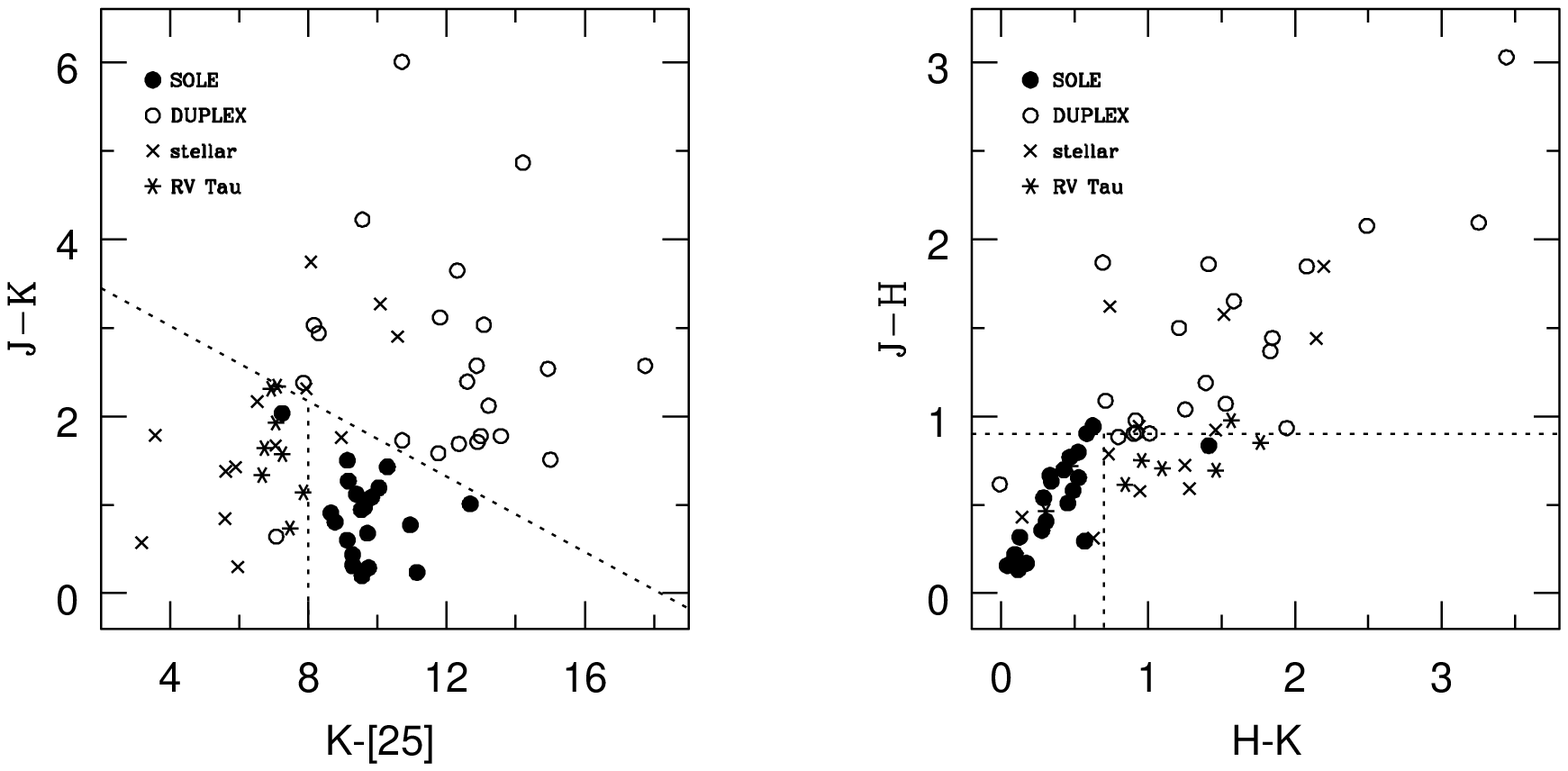}
{Fig.8. - Infrared color-color diagrams for post-AGB objects from Table 5.
Dashed lines separate parts of the diagrams occupied by different groups of
objects.} \label{ir}
\vspace{0.5cm}
\end{figure*}

The SED of IRAS 18135$-$1458 may also appear similar to the one typical for
DUPLEXes. However, while investigating this object we found that two
distinct sources separated by $\sim 5\arcsec$ were identified as this object.
One is weakly visible at optical wavelengths  with no infrared counterpart
and observed with $HST$. The other is detected only in the far-IR. The
latter is most likely the correct IRAS source (also confirmed by MSX
observations) and genuine post-AGB object. The SED for this object shown in
Figure 6 is then the composition of measurements of two different objects
and further observation should be undertaken to characterize the optical
property of this post-AGB object.

The RV Tau objects have similar SEDs with the significant infrared excess, 
but they cannot be classified according to the SED classification scheme of 
\citet{veen89}.

\subsection{Infrared color-color diagrams}
Infrared data from IRAS and 2MASS catalogues allow us to construct and
analyze infrared color-color diagrams. Fig.8 presents $J-K$ vs. $K-[25]$ and
$J-H$ vs. $H-K$ diagrams for post-AGB objects from the paper of UMB00 and
the post-AGB stars analyzed in this study. Infrared counterparts for each
star were taken from the Toru\'n catalogue of galactic post-AGB and related
objects \citep{sz07}. The transformation of 2MASS filters to Johnson system
\citep{car01} was done and IRAS flux at 25$\mu$m was converted to magnitude
by $[25]=-2.5log(F_{\nu}/6.73)$ (IRAS Explanatory Supplement 1988).

The division between different groups of objects are drawn in Fig.8.
Clustering of SOLE, DUPLEX and stellar objects at distinct locations was
already discussed by UMB00. A larger object sample in this study allowed us
to draw for the first time the specific division lines. DUPLEX sources are
the most red in $J-K$ color, and this is consistent with their optically
thicker circumstellar shell. There is also a division of DUPLEXes around
$K-[25] \sim 12$ between those with a partially visible central star
($K-[25] < 12$) and those with a fully obscured central star ($K-[25] >
12$).

SOLE objects are bluer than DUPLEXes in $J-K$ color because of a thinner 
dust envelope. The slightly shifted positions of IRAS 11385$-$5517 ($J-K$ = 
2.0, $K-[25]$ = 7.2) is caused by the prominent near-IR excess, while the 
peculiar position of DUPLEX source IRAS 09371+1212 = Frosty Leo ($J-K$ = 
0.6, $K-[25]$ = 7.1) is a result of its very bright central star and 
therefore unusual (for DUPLEX source) near-infrared colors. 

Stellar objects are the most blue in both colors, except for few peculiar
ones described in previous section. Their position on the diagram
overlapping with DUPLEXes suggest they may be also DUPLEX objects, but as we
discussed earlier, seen pole-on.

Also near-infrared diagram shows the differences between SOLE and DUPLEX
classes. SOLE objects are bluer then DUPLEX ones in both colors, with $J-H
\lesssim 1$ and $H-K \lesssim 0.7$, as it was expected for sources with
clearly visible central stars and thin dust envelopes. The position of IRAS
11385$-$5517 ($J-H$=0.8, $H-K$=1.4) is caused by a presence of hot dust.
Stellar objects are in general blue in $J-H$ color, $J-H \lesssim 0.9$,
similar to SOLE sources, but redder in $H-K$ color, $H-K \gtrsim 0.7$, more
like DUPLEX objects. It may be also a result of a hot dust very close to the
star as it is in case of 89 Her. Again, 4 stellar objects discussed in
section 4.2 are found among DUPLEXes suggesting that they may be DUPLEX
objects for which we did not detect the nebulae because of the orientation
effect.

The differences in infrared colors between morphological classes are clearly 
visible and our conclusions follow and complement the results of UMB00. 
The study of \citet{vds00} did not recognize the IR color difference,
however, there is a little overlap  between their objects and ours. This is
caused probably by selection. Their objects are selected based on IRAS
colors. They are bright in the infrared and located close to the galactic 
plane, so they are biased towards more massive objects. Some of them may not 
be post-AGB objects, e.g., IRAS 15154$-$5258 is a [WR] PN.

We also analyzed IRAS color-color diagram for the objects from our study.
The diagram is presented in Figure 9 and regions defined by \citet{veen89}
are drawn. The differences between morphological classes are also clearly
seen. SOLE and DUPLEX objects have similar [12]-[25] colors (typical for
evolved objects, located in regions IV, V and redder) but they differ in
[25]-[60] color with SOLEs being bluer and DUPLEXes being redder indicating
colder dust in DUPLEX sources. Stellar objects are bluer in both colors and 
located in regions characteristic for less evolved objects (regions III and 
VII). Four peculiar stellar objects described in the previous sections as 
possibly DUPLEX sources but seen pole-on are redder in [25]-[60] color than 
other stellar sources and located in regions VIb and IV occupied indeed by
DUPLEXes. The position of SOLE object IRAS 11385$-$5517 (in region VIII) is
due to its cold dust. $F_{12}$ for Frosty Leo is not certain (the quality
of IRAS measurement is poor) but anyway the object is very red in [25]-[60]
color (2.97) and is located outside the presented diagram. Large $F_{60}$
excess means that the object has a detached shell composed of cold matter or
a reservoir of cold matter around central star.

\begin{figure}[ht]
\includegraphics[width=\columnwidth]{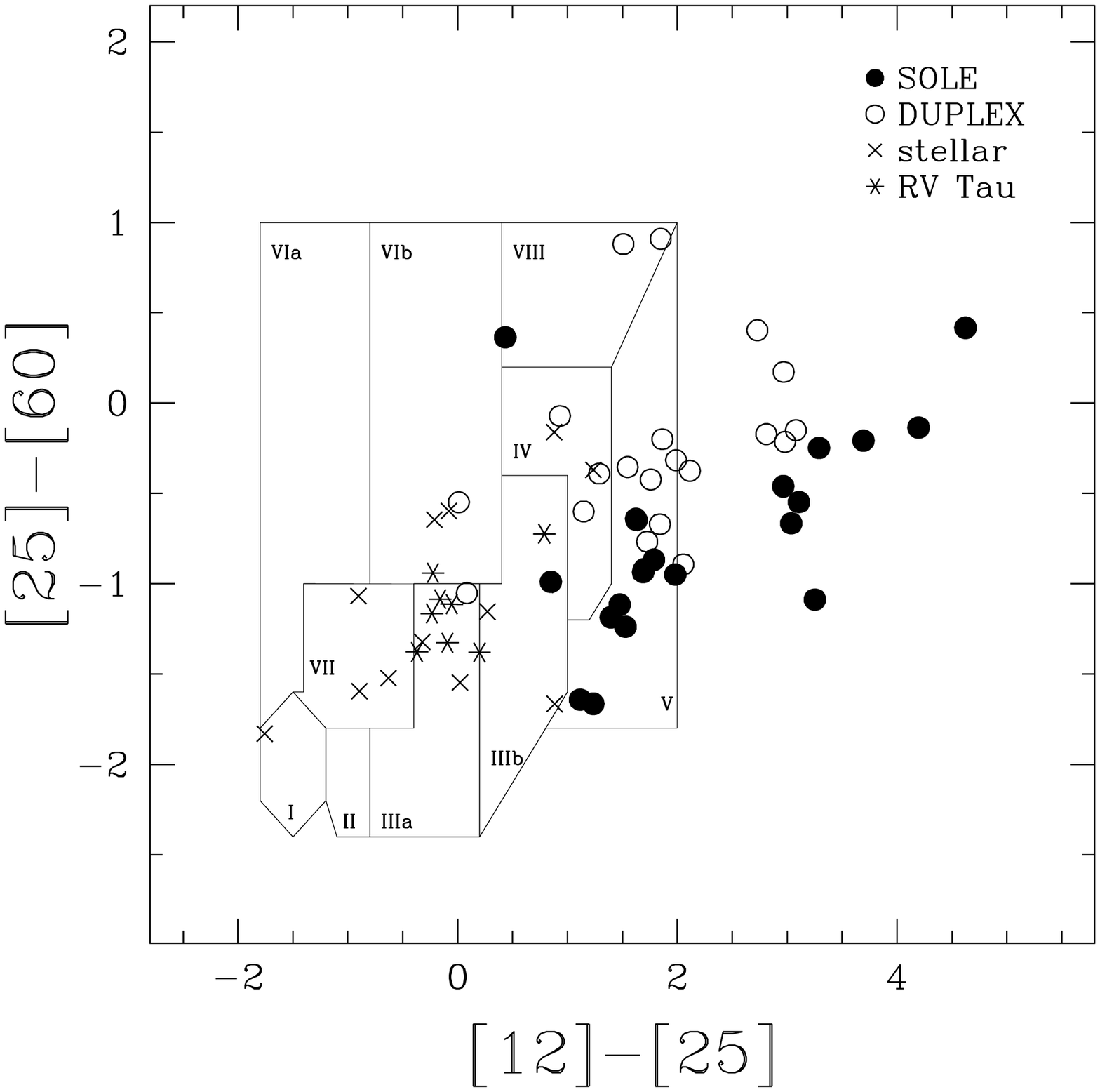}
{Fig.9. - IRAS color-color diagram for post-AGB objects from Table 5 with
regions defined by \citet{veen89}.} \label{iras}
\vspace{0.5cm}
\end{figure}

\vspace{0.5cm}

\subsection{Galactic distribution}
Galactic distribution of all post-AGB objects from Table 5 is shown in
Fig.10. DUPLEX sources are closer to the galactic plane, while SOLEs are
usually further from the galactic plane. The mean galactic latitude for
DUPLEXes is $\overline{b}_{DUPLEX}=2.35\arcdeg$ with the standard deviation
of 10.03 (however, without peculiar Frosty Leo
$\overline{b}_{DUPLEX}=1.52\arcdeg$ and standard deviation is $4.92\arcdeg$)
and for SOLEs $\overline{b}_{SOLE}=2.58\arcdeg$ with the standard deviation
of 11.23. Previous studies of PN \citep{cor95} suggested that because of
their position close to galactic plane bipolar PNs have evolved from more
massive progenitors than elliptical PNs. Following those findings, UMB00
suggested that DUPLEX objects have more massive progenitors than SOLE ones
and will evolve into bipolar PNs, while SOLEs will form elliptical PNs.
However, the masses of DUPLEX objects and some of SOLEs are not known and
this suggestion cannot be fully confirmed. The distribution of stellar
objects is quite wide with the mean galactic latitude of
$\overline{b}_{stellar}=2.94\arcdeg$ and the standard deviation of 18.07.
Some of them, like IRAS 13416$-$6243, lie very close to the galactic plane
suggesting again that they may be in fact DUPLEX sources. On the other hand,
stellar post-AGB objects far from galactic plane most likely originate from
low mass progenitors and may not become PNs.

\begin{figure}[ht]
\includegraphics[width=\columnwidth]{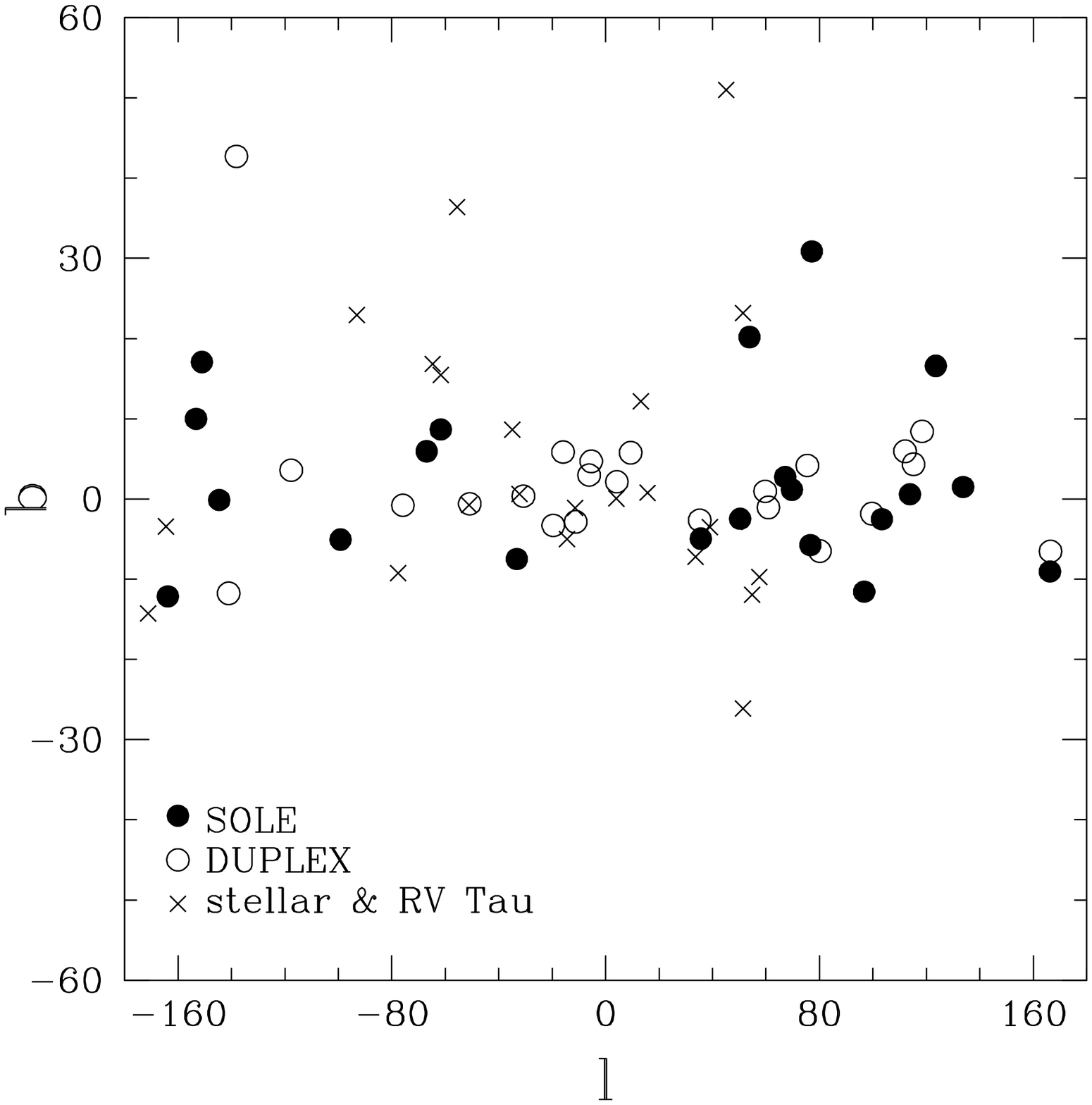}
{Fig.10. - Galactic distribution of post-AGB objects from Table 5.}
\label{galactic}
\vspace{0.5cm}
\end{figure}

\subsection{Do SOLE and DUPLEX sources have different progenitors?}
Based on previous studies of PNs and findings from observation of PPNs,
UMB00 suggested that SOLE and DUPLEX objects have different AGB progenitors
with low and high masses, respectively. More massive objects will have more
material in circumstellar shells and the star will be quite obscured, while
it won't be in a case of object with less material in the envelope. Sources
with dusty shells will therefore be fainter at the optical wavelengths but
brighter in the infrared. This was already shown on color-color diagrams in
Fig.8, with SOLE objects being bluer then DUPLEX ones. Knowing the masses of
post-AGB objects would be the best tool to confirm this dichotomy. Derived
physical parameters for 125 proto-planetary nebulae (with references) were
gathered from the literature by \citet{sta06}. Unfortunately, there are no
information for DUPLEX objects because of the lack of optical spectra from
which those parameters can be obtained. The mean mass for SOLES from this
study is $\sim 0.6M_{\odot}$, but they can be either very massive (e.g.,
IRAS 17436+5003 or IRAS 07134+1005 from UMB00 with a mass of
$M>0.8M_{\odot}$) and very ``light'' (e.g., IRAS 08143$-$4406 from this
study with a mass below 0.55M$_{\odot}$). One has to remember also about
errors on derived central star masses due to the uncertainties in effective 
temperature and gravity with mean values of $\Delta T_{\rm eff} = 350$K and
$\Delta log$g = 0.3dex (for some objects uncertainties are large and hence a
range of possible masses is big). Stellar objects in our sample are mostly
of low mass, but there are some exceptions as well (e.g., IRAS 20117+1634
with a mass of $\sim 0.9M_{\odot}$). Figure 11 shows the distribution of
masses of SOLE and stellar objects from Table 5 (masses of DUPLEXes are not
know and cannot be drawn on the diagram). Masses of stellar objects are in
general lower than masses of SOLEs, especially when we do not consider RV
Tau stars (plain columns on the diagram).

\begin{figure}[ht]
\includegraphics[width=\columnwidth]{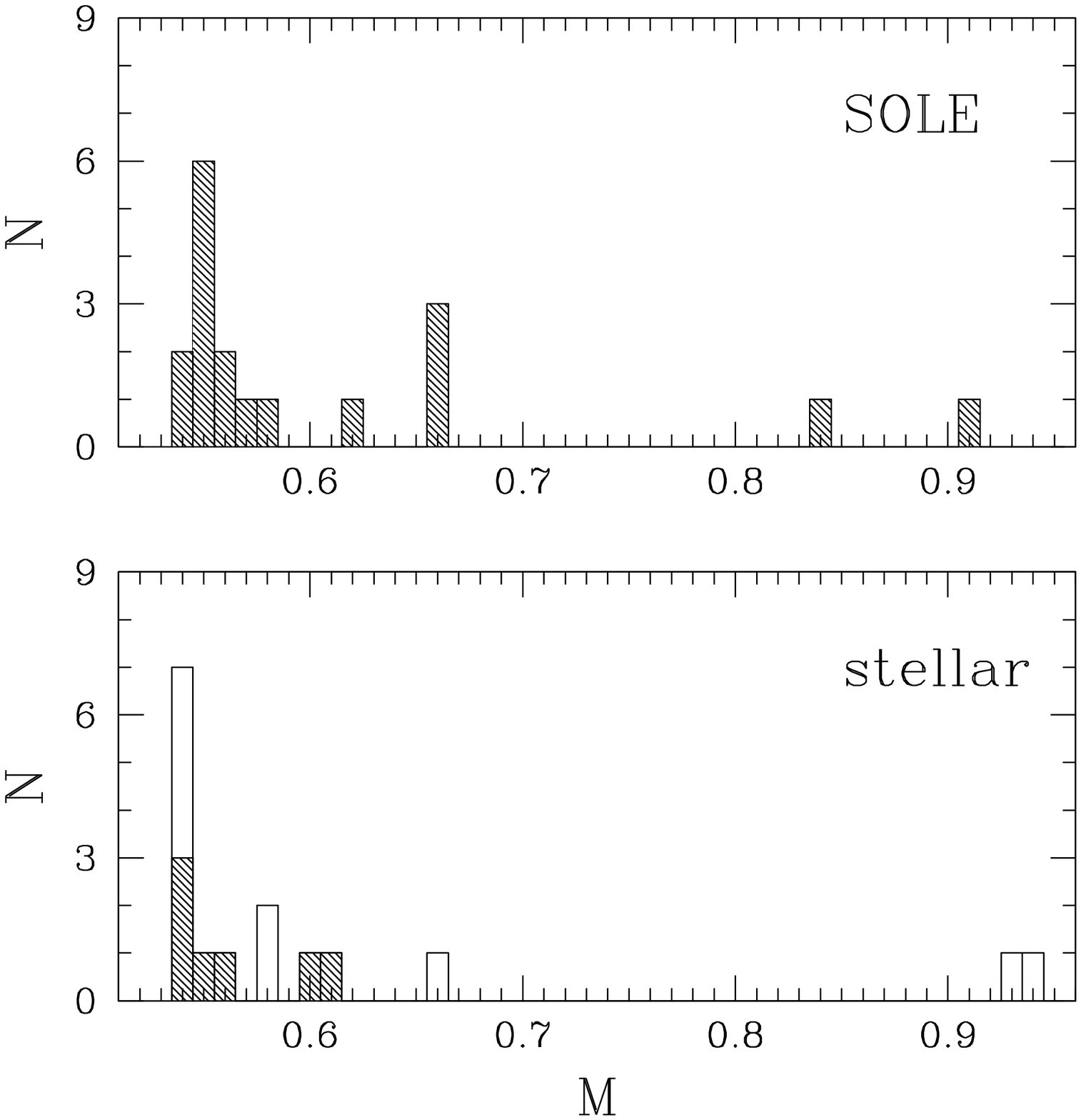}
{Fig.11. - Mass histogram for SOLEs and stellar objects from Table 5. The
most left column shows all objects with masses below $0.55M_{\odot}$. Lower
panel: plain columns correspond to RV Tau and R CrB stars.}
\label{mass}
\vspace{0.5cm}
\end{figure}

We searched also for a connection between different morphological groups and
chemistry of PPNs. In the sample of more than 50 objects from this study and
UMB00 we have 12 C-rich and 9 O-rich SOLE objects and 7 C-rich and 6 O-rich
DUPLEX objects. We see no obvious correlation between their chemistry and
PPN morphology. However, different galactic distributions of SOLEs and
DUPLEXes and therefore possibly differences in mass progenitors may suggest
also the division of chemical composition. The low mass objects do not
experience the 3rd dredge-up and evolve all the time as oxygen-rich sources.
More massive objects experience thermal pulses and their composition
undergoes significant changes, especially in carbon abundance, so they are
expected to be carbon-rich. If DUPLEXes are indeed more massive they should
be rather C-rich and SOLEs should be rather O-rich if they have indeed small
masses. However, chemical composition depends on metallicity
\citep[e.g.,][]{mar99} and stars of equal masses and different metallicity
can have different chemical abundances. Therefore the simple chemical
division in context of masses of stars is not possible. 
Since we find both O- and C-rich chemistry in both SOLEs and DUPLEXes 
in our sample, massiveness of DUPLEXes relative to SOLEs based on the shell
chemistry is not possible. However, after examining ISO, IRAS and VISIR
spectra (http://www.ncac.torun.pl/postagb and\\ http://archive.eso.org/) we
do find that all DUPLEXes with the 9.8um feature show it in absorption while
all SOLEs with the 9.8um feature show it in emission. This finding possibly
indicates that DUPLEXes tend to be optically thicker than SOLEs, suggesting
the more massive nature of DUPLEXes with respect to SOLEs. While it is an
intriguing trend, it is nevertheless based on small number statistics (only
3 SOLE: IRAS 11385-5517, IRAS 12175-5338 and IRAS 18095+2704 and 4 DUPLEX:
IRAS 08005-2356, IRAS 10197-5750, IRAS 17150-3224 and IRAS 23541+7031
objects). Thus, more observations are needed to verify this.
Most of the stellar sources in our
sample is oxygen-rich (15 O-rich and 5 C-rich) and we see no correlation
between their chemical composition and morphology and/or masses. Thus it
cannot be said that chemistry of the central star is a key leading to
differences between groups of post-AGB objects.

UMB00 found almost equal number of SOLE and DUPLEX objects, 11 and 9
respectively (the 10th DUPLEX source in their sample, IRAS 09452+1330, is an
AGB star), while in our sample there are 9 SOLE and only 4 DUPLEX post-AGBs.
The reason for this are selection criteria. The UMB00 sample focused mostly
on sources for which evidence for spatial extension existed. Our objects
were selected to complement the existing images of PPNs and to cover more
diverse properties of stars. Most of them were already studied previously
and hence we know their magnitudes, temperatures and chemistry. And usually
they are quite bright, there are not many faint objects in our sample. This
selection was also caused by requirements of snapshot survey preferring
brighter stars with shorter observation time. All this provided to the lack
of DUPLEX objects, which are usually very faint ($V\geq$15.4 in our sample).

\section{Conclusions}
We analyzed HST images of 31 post-AGB objects.  We increased the number of
observed PPNs and covered a wider variety of PPNs. Following previous
results of UMB00 we selected SOLE and DUPLEX sources based on their
morphology. Our results are consistent with UMB00, and therefore, strengthen
the validity of the general PPN structure suggested by UMB00, in which an
intrinsically axisymmetric shell assumes a different morphology mainly due
to the varying degree of the equatorial density enhancement which determines
the presence of the dust lane in DUPLEXes and the absence of it in SOLEs. We
selected a separate group of stellar post-AGB objects without visible
nebulosities. We searched for connections between morphology of the
nebulosities and the chemical and physical properties of their central stars
and did not find any obvious connection. We confirmed results of UMB00 that
SOLE objects with thin envelopes are bluer then dusty DUPLEX sources. We
also suggested that some of the stellar sources may be objects with the
nebulosity around central star, but seen pole-on and/or too faint and
distant to be resolved. We also found the differences in galactic
distribution of different groups with the SOLEs lying farther from the
galactic plane than DUPLEXes, which could suggest that DUPLEX objects may
have more massive progenitors. However, we were not able to confirm this
suggestion directly because the masses of DUPLEXes and some SOLEs are not
known.

\acknowledgments N. Si\'odmiak and M. Meixner acknowledge support from
NASA/NAG5-12595, NASA/STScI-GO-10627.01 and NASA/STScI-GO-09377.05. N.
Si\'odmiak also acknowledges support from grant N203 0661 33 of the Science
and High Education Ministry of Poland. T. Ueta acknowledges support from
STScI GO-10627.01. R. Szczerba acknowledges support from grant N203 019
31/2874 of the Science and High Education Ministry of Poland.

\begin{deluxetable}{lcccrcrl}
\tabletypesize{\small}
\tablenum{5} 
\tablecolumns{8}
\tablecaption{Morphology versus properties of post-AGB objects \label{ta}}
\tablehead{\colhead{IRAS ID} & \colhead{Prop. ID} & \colhead{SED type$^1$} &
\colhead{T$_{\rm eff}$$^2$} & \colhead{Mass} & \colhead{C/O} & \colhead{b (deg)} & \colhead{References}} 
\startdata
\multicolumn{8}{c}{\bf SOLE sources}\\
01005+7910   &10627& IVb  &21000 &     0.55 & C   &  16.59 &13,32                   \\
Z02229+6208  & 6364& IVa  & 5500 &     0.56 & C   &   1.50 &13,15,16,32,38,40       \\
04296+3429   & 6364& IVa  & 7000 &     0.55 & C   &  -9.05 &11,22,32,38,40,46       \\
05341+0852   & 6364& IVa  & 6500 &     0.55 & C   & -12.14 &13,16,32,38,46          \\
06530$-$0213 & 6364& IVa  & 7250 &     0.56 & C   &  -0.14 &11,16,20,32,38,40       \\
07134+1005   & 6737& IVb  & 7250 &     0.84 & C   &  10.00 &12,13,15,18,32,38,40,46 \\
07430+1115   & 6364& IVa  & 6000 & $<$ 0.55 & C   &  17.07 &11,16,32,38             \\
08143$-$4406 &10627& IVa  & 7150 & $<$ 0.55 & C   &  -5.07 &20,32                   \\
11385$-$5517 & 9463& IVb  & 8500 &     0.55 & O   &   5.94 &19,32                   \\
12175$-$5338 &10627& IVb  & 7350 &     0.62 & O   &   8.66 &32,45                   \\
16206$-$5956 &10627& IVb  & 8500 &     0.66 & O   &  -7.49 &7,32                    \\
17436+5003   & 6737& IVb  & 6600 &     0.91 & O   &  30.87 &9,10,16,32,38           \\
18095+2704   & 6364& IVa  & 6600 &     0.55 & O   &  20.18 &9,16,32,38              \\
19114+0002   & 6737& IVa  & 6750 &     0.66 & O   &  -4.96 &9,32,38                 \\
19306+1407   & 9463& IVa  & B    &      ... & O   &  -2.49 &13,27,48                \\
19475+3119   & 9463& IVb  & 7750 &     0.58 & O   &   2.73 &9,31,32                 \\
20000+3239   & 9463& IVa  & 5500 &      ... & C   &   1.16 &9,13,16,47              \\
20462+3416   & 6364& IVb  & B1I  &      ... & O   &  -5.75 &9,38                    \\
22223+4327   & 9463& IVb  & 6500 &     0.55 & C   & -11.56 &9,16,32,46              \\
22272+5435   & 6364& IVa  & 5750 &     0.57 & C   &  -2.52 &9,32,38,39              \\
23304+6147   & 9463& IVa  & 6750 &     0.66 & C   &   0.59 &16,32,46                \\
\multicolumn{8}{c}{\bf DUPLEX sources}\\
04395+3601   & 9430& III  &25000 &      ... & C   &  -6.53 &3,30,37                 \\
06176$-$1036 & 7297& III  & 7500 &     0.62 & C/O & -11.76 &4,5,16,32               \\
08005$-$2356 & 6364&  II  & F5   &      ... & C/O &   3.58 &16,38                   \\
09371+1212   & 9463& IVb  & K7II &      ... & ... &  42.73 &16,25                   \\
10197$-$5750 & 6816&II/III& A2I  &      ... & ... &  -0.79 &24                      \\
13428$-$6232 & 9463& III  &...   &      ... & C   &  -0.59 &41                      \\
15553$-$5230 &10627&II/III&...   &      ... & ... &   0.36 &41                      \\
16342$-$3814 & 6364& III  &...   &      ... & O   &   5.85 &16,23                   \\
16594$-$4656 & 8210& III  & B7   &      ... & C   &  -3.29 &34,35,40,42,49          \\
17106$-$3046 & 8210& III  &...   &      ... & ... &   4.70 &9,14,16,34              \\
17150$-$3224 & 6364& III  & G2   &      ... & O   &   2.98 &16,35,38                \\
17245$-$3951 & 8210& III  &...   &      ... & ... &  -2.84 &9,16,34,35              \\
17423$-$1755 & 6364&  II  & B    &      ... & O   &   5.78 &7,38                    \\
17441$-$2411 & 6364& III  & F5   &      ... & C?  &   2.15 &16,33,35,38             \\
19024+0044   & 9463& III  &...   &      ... & ... &  -2.65 &28                      \\
19374+2359   & 6364&  II  & F5   &      ... & O   &   0.96 &11,16,38                \\
19477+2401   & 8210&  II  &...   &      ... & ... &  -1.06 &9,13,16,34              \\
20028+3910   & 6364& III  & G4   &      ... & C   &   4.17 &9,16,34,38,48           \\
22036+5306   &10185&  II  &...   &      ... & O   &  -1.84 &16,26,29                \\
22574+6609   & 8210&  II  &...   &      ... & C   &   5.96 &16,34,38                \\
23321+6545   & 6364&II/III&...   &      ... & C   &   4.32 &16,38                   \\
23541+7031   & 9463&  II  & B    &      ... & O   &   8.42 &2,18                    \\
Egg Nebula   & 9463&II/III& 6500 & $>$ 0.94 & C/O &  -6.50 &21,32                   \\
\multicolumn{8}{c}{\bf Stellar sources}\\
05113+1347   & 6364& IVa  & 5250 &     0.60 & C   & -14.29 &16,32,38                \\
10158$-$2844 & 6364& N/A  & 7600 &     0.56 & C/O &  22.94 &5,16,32,38              \\
12222$-$4652 &10627& N/A  & 6800 & $<$ 0.55 & O   & -15.49 &5,32,45                 \\
12538$-$2611 &10627& N/A  & 6000 & $<$ 0.55 & O   &  36.40 &32,45                   \\
13416$-$6243 &10627&  II  &...   &      ... & C   &  -0.73 &36,41                   \\
15039$-$4806 &10627& N/A  & 8000 &     0.55 & O   &   8.65 &32,45                   \\
17163$-$3907 &10185&  II  &...   &      ... & ... &  -1.12 &18                      \\
17279$-$1119 &10627& N/A  & 7300 & $<$ 0.55 & C   &  12.17 &32,45                   \\
17516$-$2525 & 9463&  II  &...   &      ... & O   &   0.06 &36,44                   \\
17534+2603   &10627& N/A  & 6550 &     0.61 & O   &  23.19 &5,16,32,50              \\
18135$-$1456 & 9463& III  &...   &      ... & O   &   0.77 &6,43                    \\
20136+1309   & 8210& IVa  & G0   &      ... & ... & -11.97 &34                      \\
20547+0247   & 9463&  II  & G    &      ... & O   & -26.11 &1,8                     \\
RV Tau stars:\\
06034+1354   &10627& N/A  & 5900 & $<$ 0.55 & O   &  -3.42 &5,32                    \\
09256$-$6324 &10627& N/A  & 6700 & $<$ 0.55 & C   &  -9.24 &5,32                    \\
12067$-$4508 &10627& N/A  & 6000 & $<$ 0.55 & O   &  16.82 &5,32                    \\
15469$-$5311 &10627& N/A  & 7500 & $<$ 0.55 & O   &   0.63 &5,32                    \\
17243$-$4348 &10627& N/A  & 6750 &     0.94 & O   &  -4.99 &5,32                    \\
19125+0343   &10627& N/A  & 7750 &     0.58 & O   &  -3.49 &5,32                    \\
19157$-$0247 &10627& N/A  & 7750 &     0.58 & O   &  -7.23 &5,32                    \\
20117+1634   &10627& N/A  & 5000 &     0.93 & O   &  -9.76 &5,32                    \\
R CrB stars:\\
15465+2818   & 6364& N/A  & 6705 &     0.66 & C   &  50.98 &16,32,38                \\
\enddata
\end{deluxetable}

$^1$ Types of spectral energy distribution according to \citet{veen89}. SEDs that cannot be classified are marked as N/A.

$^2$ Spectral type is shown if T$_{\rm eff}$ is not known from the model atmosphere analysis.

\vspace{0.5cm}

REFERENCES: All objects are included in the  Toru\'n catalogue of galactic
post-AGB and related objects \citep[][http://www.ncac.torun.pl/postagb]{sz07}.
1.~\citet{bar96}, 2.~\citet{cas02}, 3.~\citet{cer01},
4.~\citet{coh04}, 5.~\citet{ruy06}, 6.~\citet{eng02}, 7.~\citet{gau04},
8.~\citet{geb05}, 9.~\citet{gle01}, 10.~\citet{gle03}, 11.~\citet{gle05},
12.~\citet{hony03}, 13.~\citet{hriv00}, 14.~\citet{kwok00},
15.~\citet{kwok02}, 16.~\citet{mei99}, 17.~\citet{mei04}, 18.~\citet{nym98},
19.~\citet{olo99}, 20.~\citet{rey04}, 21.~\citet{sah98}, 22.~\citet{sah99a},
23.~\citet{sah99}, 24.~\citet{sah99b}, 25.~\citet{sah00}, 26.~\citet{sah03},
27.~\citet{sah04}, 28.~\citet{sah05}, 29.~\citet{sah06}, 30.~\citet{san02},
31.~\citet{sar06}, 32.~\citet{sta06}, 33.~\citet{su98}, 34.~\citet{su01},
35.~\citet{su03}, 36.~\citet{sz03}, 37.~\citet{tram02}, 38.~\citet{ueta00},
39.~\citet{ueta01}, 40.~\citet{ueta05}, 41.~\citet{vds00}, 42.~\citet{vds03},
43.~\citet{veen89}, 44.~\citet{veen89a}, 45.~\citet{win97}, 46.~\citet{win00a},
47.~\citet{volk02}, 48.~\citet{volk04}, 49.~\citet{volk06},
50.~\citet{wat93}.

\end{document}